\documentclass[twocolumn,aps,showpacs,prb,epsf,graphics,psfig]{revtex4-2}
\usepackage{graphicx}
\usepackage{graphicx}
\usepackage{bbold}
\usepackage{color}
\usepackage[normalem]{ulem}
\usepackage{amssymb}
\usepackage{tikz}
\usepackage[version=3]{mhchem} % \ce{^{.}OH}-radicals
\usepackage{amsmath}
\begin{document}
\title{A molecular dynamics simulation of the abrupt changes in the thermodynamic properties of water after formation of nano-bubbles / nano-cavities induced by passage of charged particles}
\author{Ramin Abolfath$^{1,5\dagger}$, Niayesh Afshordi$^{2,3,4}$, Sohrab Rahvar$^{5}$, Adri van Duin$^{6}$, Martin R\"adler$^{7}$, Reza Taleei$^{8}$, Katia Parodi$^{7}$, Julie Lascaud$^{7}$, Radhe Mohan$^1$}
\affiliation{
$^1$Department of Radiation Physics and Oncology, University of Texas MD Anderson Cancer Center, Houston, TX, 75031, USA \\
$^2$Department of Physics and Astronomy, University of Waterloo, Waterloo, ON, N2L 3G1, Canada \\
$^3$Waterloo Centre for Astrophysics, University of Waterloo, Waterloo, ON, N2L 3G1, Canada\\
$^4$Perimeter Institute for Theoretical Physics, 31 Caroline St. N., Waterloo, ON, N2L 2Y5, Canada \\
$^5$Physics Department, Sharif University of Technology, P.O.Box 11365-9161, Azadi Avenue, Tehran, Iran \\
$^6$Department of Mechanical and Nuclear Engineering, Pennsylvania State University, Pennsylvania 16802, USA \\
$^7$Department of Medical Physics,  Ludwig-Maximilians-Universit\"{a}t M\"{u}nchen, Munich, Germany \\
$^8$Department of Radiation Oncology, Jefferson University, Philadelphia PA, USA
%$^8$ Department of Radiation Oncology, Washington University School of Medicine in St. Louis, St. Louis, Missouri 63110, USA
}

\date{\today}
%%%%%%%%%%%%%%%%%%%%%%%%%%%%%%%%%%%%%%%%%%%%%%%%%%%%%%%%%%%%%%%%%%%%%%%%%%
\begin{abstract}
{\bf Purpose}:
To present a multi-scale formalism that accounts for the formation of nano-scale bubbles/cavities owing to a burst of water molecules after the passage of high energy charged particles that leads to the formation of ``hot" non-ionizing excitations or thermal spikes (TS).

{\bf Methods}:
We constructed amorphous track structures to account for the formation of TS in a step-by-step Monte Carlo (MC) simulation of ionizing radiation in liquid water.
%to score the spatial and temporal distribution of energy depositions.
Subsequently, we simulate sudden expansion and collective motion of water molecules employing a molecular dynamics (MD) simulation that allows computation of ${\cal O}(10^6)$ particle trajectories and breaking/forming of chemical bonds on the fly using a reactive force field, ReaxFF.
We calculate the fluctuations of thermodynamic variables before and after TS formation to model the macroscopic abrupt changes of the system, possibly the occurrence of a first-order phase transition, and go beyond the accessible simulation times by engaging fluid dynamic equations with appropriate underlying symmetries and boundary conditions.
%We cross-validate the time-evolution of water molecules obtained from MD and go beyond the accessible simulation times by engaging Navier-Stokes equations in fluid dynamics using appropriate underlying symmetries and boundary conditions.

{\bf Results}:
We demonstrate the coexistence of a rapidly growing condensed state of water and a hot spot that forms a stable state of diluted water at high temperatures and pressures, possibly at a supercritical phase.
Depending on the temperature of TS, the thin shell of a highly dense state of water grows by three to five times the speed of sound in water, forming a thin layer of shock wave (SW) buffer, wrapping around the nano-scale cylindrical symmetric cavity.
The cavity contains a low density of water molecules; hence, it forms a bubble-like state of water embedded in a room-temperature background of water.
The stability of the cavity as a result of the incompressibility of water at ambient conditions and the surface tension allows the transition of supersonic SW to a subsonic contact discontinuity and dissipation to thermo-acoustic sound waves.
Thus, TS gradually decays to acoustic waves, a channel of deexcitation that competes with the spontaneous emission of photons, and a direct mechanism for water luminescence. %scintillation.
We further study the mergers of nanobubbles that lead to fountain-like or jet-flow structures at the collision interface.
We introduce a time delay in the nucleation of nano-bubbles, a novel mechanism, responsible for the growth and stability of much larger or even micro-bubbles, possibly relevant to FLASH ultra-high dose rate (UHDR).

{\bf Conclusions}:
The formation of fast-growing nano-bubbles is an indication of a much higher diffusion constant currently implemented in the reaction-diffusion of the Monte Carlo models in the radiation physics community.
The current study is potentially significant in modeling overlap integrals and inter-track recombination processes among the radiation tracks at UHDRs and is useful for the FLASH radiotherapy data analysis.
Our analysis predicts that thermal radiation from the transient supercritical state of water localized in nano-cavities wrapping around the track of charged particles can be manifested in the (indirect) water luminescence spectrum.
\end{abstract}
%\pacs{87.50.-a, 87.53.-j, 87.55.N-}
\pacs{}
\maketitle
%%%%%%%%%%%%%%%%%%%%%%%%%%%%%%%%%%%%%%%%%%%%%%%%%%%%%%%%%%%%%%%%%%%%%%%%%%
\section{Introduction}
The formation of microscopic bubbles [\onlinecite{Norman1963:NSE,Vedadi2010:PRL,Min2019:JCP}] with the size of sub-microns in water has attracted interest due to a potential paradigm shift in the treatment of cancer, such as hypothermia [\onlinecite{Mallory2016:CROH}], and sonodynamic therapy [\onlinecite{Wan2016:CBM}].
Understanding a potential connection with other recently developed treatment techniques, such as spatially fractionated radiotherapy [\onlinecite{Yan2020:CTRO}], and FLASH radiotherapy [\onlinecite{Favaudon2014:STM,Montay-Gruel2018:RO,Vozenin2018:CCR,Montay-Gruel2019:PNAS,Buonanno2019:RO,Vozenin2019:RO,Arash2020:MP,Spitz2019:RO,Koch2019:RO,Abolfath2020:MP,Seco2021:MP,Abolfath2023:FP,Baikalov2023:FP,Oraiqat2020:MP}], may impact the clinical applications of these modalities. However, the topic is more directly relevant to various imaging techniques designed for the detection of charged particle tracks by thermo-acoustic [\onlinecite{Kalinichenko2001:chapterbook,Oraiqat2020:MP,Hickling2018:MP,Lascaud2023:PMB}], optical (scintillation) [\onlinecite{Beddar2016:Book}] and nuclear magnetic resonance (NMR) beam topography or, by magnetic resonance imaging (MRI) [\onlinecite{Odeen2019:PNMRS,Gantza2023:PNAS}],
including a long list of other scientific and engineering applications from nuclear physics [\onlinecite{Winter2020:ANE}] to cold fusion and sonoluminescence effect [\onlinecite{Brenner2002:RMP}], astrophysics, high-energy physics and space radiation [\onlinecite{Durante2011:RMP}].
% states of water in extreme conditions.
%In particular, micro-bubbles formed by focusing the low-intensity ultrasound beams in sonodynamic therapy (SDT) has emerged as a new treatment modality with the capability to treat deep-seated tumors without invasion and has overcome limited penetration of visible light beams in photodynamic therapy.

The passage of ionizing radiation results in the formation of ionizing and non-ionizing energy deposition [\onlinecite{{Abolfath2022:PMB}}] to molecules within a nanometer radial distance from the center of the track.
Non-ionizing excitations deposited into such a cloud of molecules release their internal energy either in a slow nanosecond process by spontaneous emission of photons or in a fast femtosecond process through the host nuclear vibrations. The latter further transforms the release of excitation energies into heat exchange with the surrounding molecules.
Statistically, the spatial dimension of these cylindrical symmetric clouds of molecules averaged over a large ensemble of tracks, is a rapidly decreasing function of distance from the center of the track, $b$. See Eqs. (\ref{eq25r}) and (\ref{eq24x}), analogous to a quadratic falling ($1/b^2$) of the dose profile  shown in [\onlinecite{Wang2014:PMB}].
Under such an ensemble average, the spatial details of individual tracks that exhibit the inchoate distribution of energy transfer [\onlinecite{Kellerer1985:chapterbook}] are washed out, converging into an amorphous track structure.

Although for several decades this energy transfer has been known to lead to the formation of thermal spikes (TS) [\onlinecite{LaVerne2000:RR}] and microscopic gas bubbles in liquids [\onlinecite{Norman1963:NSE}], the mechanism and its possible role in advanced radiotherapy applications have not been fully understood. In particular, TS's wrap around the cold ions in the medium, boosting the initial diffusion of species. This acts as a driver for the outward burst of ions, and consequently, it can potentially impact the dynamics of the fast chemical reaction following ionizing radiation [\onlinecite{Abolfath2022:PMB}]. In radio/particle therapy, fast chemical and early biological processes can be further influenced by the shock waves emerging from TS [\onlinecite{Surdutovich2014:EPJD}], as well as bubble expansion and collapse [\onlinecite{deVera2018:EPJD}]. These phenomena may be of greater importance for treatment outcomes in novel radiotherapy schemes such as FLASH radiotherapy, in which the dose is administered at an ultrahigh dose rate (UHDR $\geq$ 40 Gy/s) in less than a second to reduce toxicity in healthy tissues [\onlinecite{Favaudon2014:STM,Montay-Gruel2018:RO,Vozenin2018:CCR,Montay-Gruel2019:PNAS,Arash2020:MP,Koch2019:RO}]. In these conditions of ultra-high fluence, the average distance between particle tracks is such that they are close enough to interact. This has already been shown to influence ultrafast chemical processes and is assumed to be one of the main reasons for the decrease in toxicity in FLASH-RT [\onlinecite{Abolfath2023:FP}]. However, current simulation frameworks fail to reproduce experimental observations, highlighting the need for more accurate modeling engines that account for physical phenomena neglected in current state-of-the-art models.

%In particular, recent efforts on developing of ionizing radiation acoustic imaging (iRAI) technique for real-time dosimetric measurements for FLASH radiotherapy [\onlinecite{Oraiqat2020:MP}] may open up an in-depth understanding of the first principle mechanisms consistent with the picture explored in this work, i.e. that they start from the formation of TSs and end with an acoustic pulse.

%The theoretical modelings and studies of physico-chemical processes start from the enumeration of the stochastic energy deposition in water, cells, and tissues. It is known that fast moving particles transfer their kinetic energy to medium and generate ionizing and non-ionizing excitations. The latter is a source of thermal spikes distributed radially around primary charged particles and form amorphous track structures.

Monte Carlo (MC) samplings of chemical species generated by radiation tracks are the state-of-the-art computational model for the analysis and computer simulations of the physico-chemical processes [\onlinecite{Agostinelli2003:NIMA,Incerti2010:IJMSSC}].
The latest version of these toolkits has added a series of chemical reaction rates based on empirical data at low dose rates of homogeneous $\gamma$-rays.
They, however, do not completely model the collective dynamics of reactive species and change of medium, the processes relevant to interplay between change in the heterogeneous chemical changes of the tracks and the dynamics of TS's at nano-scales.

As noted above, current MC models cannot account for the FLASH-UHDR experimental observations. Thus, they need substantial refinements to become reliable computational tools for planning FLASH-UHDR radiotherapy.
Our technical aim is to go beyond this simple framework by simulating the evolution and interaction of TS's, produced in the wake of particle tracks, using large-scale molecular dynamics (MD) at UHDR to demonstrate the collective flow of molecules and introduce a mechanism that constitutes elevation of reaction and diffusion constants of ionized species.

In this work, we present our theoretical predictions of the formation of nano-bubbles, pico-seconds after generation of TS's.
Our motivation in this work is to present the MD-based computational tool for the characterization of the nano-bubbles and investigate the interplay between the TS's and the inter-track correlations such as recombination processes of chemical species that are potentially substantial for the radio-biological effects and translational applications of FLASH radiotherapy and their predictive outcomes.

%Because of their highly intensive energy stored within a nano-meter scale, they spark shock-waves that form a moving high pressure outward. That pushes water molecules outside of the track and leaves a low pressure with low water density behind SW. We call this a nano-bubble or a nano-cavity.

Our calculation shows a rapid evolution from shock to contact discontinuity after the creation of a thermal spike, where the pressure gradients inside and outside the bubble cancel each other out. They act at two different locations (inside and outside of the bubble interface) with opposite gradients with zero curvature at the interface, which makes the bubble dynamically stable. Any fluctuation inside the bubble that causes an increase in the negative slope of the pressure (making it more negative) breaks the symmetry/balance in the pressure profile and drags the molecules toward the center of the bubble (like creating a vacuum in the center), which eventually enforces the bubble to collapse suddenly.
%This is similar to a sonoluminescence effect but at the nano-scale.

%I am looking for a model that couples the pressure profile with a fluctuating field. In the literature on Casmir effect and sonoluminescence, people talk about the dipole moments and coupling with zero-point fluctuations of EM fields (ideally at zero temperature) but are they strong enough to change the direction of the pressure gradient significantly and collapse the bubble? By the way, the temperature inside the bubble is very high. Any insight on this would be highly appreciated.

%%%%%%%%%%%%%%%%%%%%%%%%%%%%%%%%%%%%%%%%%%%%%%%%%%%%%%%%%%%%%%%%%%%%%%%%%%
%%%%%%%%%%%%%%%%%%%%%%%%%%%%%%%%%%%%%%%%%%%%%%%%%%%%%%%%%%%%%%%%%%%%%%%%%%

\section{Method}
Our methods to investigate the effects of TS's and their thermal-diffusion processes in a medium consist of two parts:
(1) molecular dynamics (MD) simulations using LAMMPS-ReaxFF with geometries and symmetries obtained from amorphous track structure, and
(2) solving a set of non-linear fluid dynamics partial differential equations with appropriate boundary conditions,
a coarse-grained continuous model.
%To simplify the computation, we calculate the propagation of the waves and fluid motion in a radial direction assuming a cylindrical symmetric track structure, consistent with the geometry of amorphous track-structures.
%In the second part, we cross validate the predictions at the microscopic scale by calculating trajectories of approximately half of a million water molecules with 0.1 fs time-steps using molecular dynamics simulations with ReaxFF.

\subsection{Amorphous track structure calculation}
A model based on the amorphous track structure calculated by MC or other analytical formulations [\onlinecite{Fain1974:RR,Wang2014:PMB}] allows the construction of initial conditions in MD that start from the following assumptions:
(1) sub-femtosecond passage of a charged particle (e.g., proton) through nm-thick MD slab,
(2) instantaneous propagation of electromagnetic (EM)/photon fields to reach the nm-size edge of a cylindrical symmetric TS,
(3) sparse and relatively slow delta-rays crossing out the MD box thus we neglect their effects in forming TS inside of our simulation box.
Violating these assumptions does not limit us to constructing the initial condition in MD but reduces complex initial conditions to simplistic ones.
For ease of calculation and proof of principle, we chose to present our results based on these simple initial conditions.

%We neglect the effects of delta-rays in the creation of TS's because the charged-particle fields create dense molecular excitations instantaneously in contrast to slow-moving and highly sparse energy depositions by delta-rays.
%Note that 60\% - 70\% of energy deposition by the charged particles is spent on the creation of ions and delta-rays.

%However because the number of delta-rays is sparse and they travel much longer beyond the nm-scale size of our computational box, from the position of the charged particles,
%we can neglect their contribution to temperature increase and molecular thermal excitations surrounding the charged particles.
%We can however start from a more complex distribution of the thermal energies among atoms, but such simplifications used in starting from a uniform distribution do not lower the generalities of our propositions, results, and conclusion.

We neglect the contribution of the delta-rays with a range beyond the nm lateral dimension of the TS confined in an MD computational box. They do not contribute significantly to temperature increases and molecular thermal excitations surrounding the charged particles. Note that 60\% - 70\% of the energy deposition by the charged particles is spent on the creation of ions and delta-rays. However, delta-rays originating from the position of the high energy (low LET) charged particles travel far beyond the nm-scale size of our computational box. In contrast, the delta-rays generated from low-energy (high LET) charged particles carry low kinetic energies (KE) but with high numbers and compactness. They deposit their energies within nm range from the position they were created. Therefore, in a first assumption, the charged-particle fields instantaneously create much denser molecular excitations than the slow-moving and highly sparse energy deposition by delta-rays. The considered uniform distribution does not lower the generalities of our propositions, which could later integrate a more complex distribution of the thermal energies among atoms.

In nanometer scales, relevant to our MD simulation geometries, the instantaneous energy deposition within the subfemtosecond time scale, and the time frame for sequential steps in MD is justifiable.
To illustrate, let us consider a fast proton passing through a 1 nm thick water slab in 0.1 fs, the spatial and temporal scales in our MD simulations.
To fulfill this condition, the charged particle is required to pass through with minimum kinetic energy calculated by
$KE = (\gamma_L - 1)M c^2$ where $\gamma_L = (1-\beta^2)^{1/2}$ and $\beta=v/c$. For a proton with a rest mass of $M = 938.272$, MeV this requires $KE \geq$ 0.52 MeV.
For simplicity in calculating this threshold in proton KE, we disregarded the deceleration of the proton along 1 nm distance and the retardation of the electric field for the energy transform to that volume.
Taking into account these factors, we may end up with a slightly higher KE threshold.
Note that this energy is much lower than the nominal energy of protons used in radiotherapy, i.e. 70 to 230 MeV.
Considering the range of protons as a function of KE, only protons at the very end of the track violate our condition for generating cylindrical symmetric TS.
In amorphous track structure approximation protons below these minimum kinetic energies tend to generate conical TS's with circular symmetries, i.e., a graded temperature profile.
However, the rise in temperature for such low-energy particles is not significant.
We finally note that the energy transfer in the lateral dimension takes place through the propagation of EM fields below the time scales 0.001 to 0.01 fs, assuming a uniform spread of energy transfer in a cylinder with a 1 nm diameter.
Thus, the formation of molecular excitations due to electron-proton impulsive Coulomb interaction is much faster than the MD time step (0.1 fs or 100 as).
Therefore, it would be a good approximation to consider molecular excitations as an initial condition for the simulation of the TS propagation.

Fig. \ref{Fig1} illustrates a schematic geometry of the amorphous track structure considered in this work.
The circles in Fig. \ref{Fig1} represent the radius of isoline energy deposition of such amorphous tracks in our MD simulations.
%However, in Geant4-DNA formation of chemicals scores above 1 ps time steps, which corresponds to the time scale of delta-rays. Thus we are suspicious in Geant4-DNA, the formation of chemical species by molecular excitations, and thermal spikes, within a 1 nm radius in the vicinity of proton tracks is not taken into account.
\subsection{Molecular dynamics}
% https://docs.lammps.org/velocity.html
We employ a million particles MD simulation to calculate the trajectories of water molecules, using LAMMPS.
We chose ReaxFF as a model for inter-particle interactions to be able to capture the molecular dissociation including breaking and forming chemical bonds during the time evolution of particles.

ReaxFF [\onlinecite{vanDuin2001:JPCA}] is a general bond-order-dependent potential that provides accurate descriptions of bond breaking and bond formation.
It is a bond-order-dependent empirical force field method, which includes a polarizable charge function, enabling application to a wide range of materials and accurate reproduction of reaction energies and barriers.
Recent simulations on a number of hydrocarbon-oxygen systems [\onlinecite{Yusupov2012:NJP,Verlackt2015:NJP}] and DNA [\onlinecite{Abolfath2011:JPC}] showed that ReaxFF reliably provides energies, transition states, reaction pathways, and reactivity trends in agreement with QM calculations and experiments.

For all the simulation results presented thereafter, the MD time step is constant for a given run and chosen to be within 0.1-0.25 fs depending on the desired total simulation time and temporal resolution. Time steps greater than these values lead to numerical instability and early termination of the simulation. Therefore, the physical simulation time can be easily calculated by multiplying the total steps and the time step.

Information on the type of chemical reactions and the time evolution of damage is collected by running the MD up
to 50 ps, where the rearrangement of the atomic coordinates has been deduced from a dynamical trajectory calculated by ReaxFF.
These simulations were performed using periodic and/or free boundary conditions in canonical moles, pressure, and temperature (NPT), as well as a constant energy and volume (NVE) ensemble with a Nose-Hoover thermostat for temperature control.

In MD, NPT refers to a system of particles, equivalent to a canonical ensemble, with a constraint on the values of the parameters, N, P, and T. Practically a thermostat with virtual/auxiliary degrees of freedom is added to apply a fluctuation-dissipation dynamics to the entire system, such as coupling with a thermal bath following a series of Langevin equations to enforce convergence of the thermodynamic parameters to the user-defined values. Because of the constant P, the volume of the system, V, is a variable in time.
Similarly, for NVE the constraints apply to constant parameters V and total energy E. Here, P is a free parameter and a time-dependent variable. The system of particles as such is equivalent to a microcanonical ensemble.

We initially allow the entire system to run for several thousand steps to reach thermal equilibrium at room temperature.
Once the fluctuations in temperature and pressure are damped, the passage of a charged particle and induction of
all ionizations and excitations scored within the MD time step (e.g., 0.1 fs) that separates the state of the system before and after the passage of fast-charged particles were performed.
In amorphous track structure construction, the ionizations and production of secondary electrons and delta rays are neglected as they are sparse in their numbers and dimensions, reaching outside of the core of TS domains.
In contrast, the effects of non-ionizing excitations are lumped and integrated into an abrupt change in the velocities of a subset of water molecules forming TS's.
These new velocities can be considered an initial condition for the rest of the simulation.
We then allow the system to evolve spontaneously and capture the trajectories of the millions of particles.

Using the create style in LAMMPS, we set an initial condition to add TS's by changing the velocities of a group of atoms in a cylinder.
The water molecules outside the cylinder have a spectrum of kinetic energies that fluctuate thermally around room temperature and ambient pressure with a macroscopic mass density of room-temperature water (997 kg / m$^3$).
This method provides an ensemble of velocities from a uniform distribution, using a random number generator with the specified seed at the specified temperature.
It selects velocities randomly within a range of minimum to maximum values and subsequently scales the velocities to match kinetic energies to the assigned temperature.
In our setup, the updated velocities are the sum of new and old velocities (calculated under equilibrium conditions at room temperature).
The total velocities are subjected to two constraints to force the system of water molecules in the TS to contain zero total linear and angular momentum.

\subsection{MD computational box}
%The radius of cylindrical TS's was extracted from an amorphous track structure calculation using Geant4-DNA, with a mean value of the radius $\approx 1$ nm.

The radius of cylindrical TS's can be extracted from an amorphous track structure calculation using MC toolkits such as Geant4 and Geant4-DNA [\onlinecite{Agostinelli2003:NIMA,Incerti2010:IJMSSC}] or using analytical formulation [\onlinecite{Wang2014:PMB}]. Hereby, we consider a cylinder with a mean radius $\approx 1$ nm based on previous works [\onlinecite{deVera2019:CN,deVera2018:EPJD,Abolfath2022:PMB}].

Considering a single track was penetrating in an MD computational box. The lateral and longitudinal size of the box can be determined by the radiation deposited dose in Gy (J/kg) and the charged particle LET in eV/nm (keV/$\mu$m), respectively.

For an illustration of the problem, let us consider a single track of $^{12}$C close to its Bragg peak with LET of $8.45\times 10^2$ eV/nm creating a cylindrical track with a radius of 1 nm and a length of 4 nm.
The mass of water in this cylinder, $m=\rho_m V = 13\times 10^{-21}$ g, where $\rho_m=$1 g/cm$^3$ and $V=\pi(1nm)^2\times 4(nm) \approx 13 nm^3$ is the volume of the cylinder.
Considering that the energy deposited by $^{12}$C uniformly distributed along the axis of the cylinder, we find $\Delta E = {\rm LET} \times L = (8.45\times 10^2 {\rm eV/nm})\times 4 {\rm nm} = 34 \times 10^2 eV = 5408 \times 10^{-19} J$.
The deposited dose to this volume is then $D = 5408 \times 10^{-19} J/(13\times 10^{-24} kg) = 41.6\times 10^6$ Gy.

This is, of course, a very large dose, that is, 6 to 8 orders of magnitude larger than the radiation dose in medical applications.
However, by extending the volume to an entire MD computational box, and including the mass of water molecules in the denominator of $D$, the numerical value of the dose relative to the calculated dose deposited in the cylinder is substantially lower.
The same is true for the increase in temperature, $\Delta T$, using the water-specific heat, $C_V$, where the increase in temperature throughout the MD computational box is substantially lower than $\Delta T$ calculated only within the cylinder.

In this type of calculation, one needs to remove high-energy delta-rays and any secondary particles that deposit their energies outside of the MD box. However, we consider low-energy secondary electrons. The ones deposit energy in nm distance from the primary particle.
For simplicity in our numerical illustration, because the majority of the secondary electrons in low energies are generated close to the Bragg peak of a primary charged particle, we simply do not take into account an escape of energy from the MD box due to delta-rays.

As the deposited dose scales down linearly by increasing the lateral dimension of the computational box (considering the entire mass of water in the box for the calculation of the mass, in the denominator of the dose), we tend to consider an MD box with a size of $10^3\times 10^3 {\rm nm}^2 (1\times 1 \mu m^2)$, or even $10^4\times 10^4 {\rm nm}^2 (10\times 10 \mu m^2)$. The latter corresponds to a uniform lateral distribution of the single tracks per single-cell nucleus.

In MD, using ReaxFF, a computational box with sizes of $10^3\times 10^3 {\rm nm}^2$ and $10^4\times 10^4 {\rm nm}^2$ is considered very large.
To put this in perspective, consider a box with lateral and longitudinal dimensions of $8\times 8 {\rm nm}^2$ and 4 nm.
We can fit this box with 24,399 hydrogen and oxygen atoms to construct 8,133 H$_2$O molecules with an average mass density of water under ambient conditions (1 g/cm$^3$). To scale up this number to a system of water molecules in a box of $8,000\times 8,000~{\rm nm}^2$ we need to nearly scale up 24,399 atoms to 24,399 million atoms, which is a huge number of particles.
This is beyond current computational capabilities.

To be able to perform MD simulations, we, therefore, need to limit ourselves to smaller dimensions and truncate the number of particles up to a few million fitted to up to $100\times 100~{\rm nm}^2$ computational box, which is possibly the largest dimensions our current brute-force computational platform can handle.

By truncating the number of atoms, we may encounter some computational artifacts such as the rate of temperature decay of the entire system or the rate of expansion of the TS wavefront (using periodic boundary conditions) that we will discuss in detail below.

%In the future, we may be able to scale up our computation to billions of atoms by adopting AI/ML algorithms to MD simulations, hence removing the computational limitations and the artifacts for medical applications of ionizing radiation.

Note that for a given dose deposited by lighter charged particles than $^{12}$C, such as the proton and electron, the minimum size of the computational box is significantly smaller than the one we estimated above. This value and the temperature of the core of TS scale up like $Z^2$ where $Z$ is the charged particle atomic number.

\section{Results}
Our results consist of the derivation of analytical equations, numerical calculations, and MD simulations as follows.
\subsection{Thermal-spike formation}
We divide energy deposition into ionization and molecular excitation.
The molecular excitations that constitute TS are localized in space, within nm distance from the core of the track where the charged particle passes through.
The charged particle induces an impulse to valence electrons that transfers to their host nuclei due to electron-nucleus interaction that in turn converts to the excess in the nuclear recoil kinetic energy, a mechanism that is faster by at least five orders of magnitude than spontaneous emission of optical photons [\onlinecite{Ramin_Martin:Unpublished}].
Thus, it is plausible to assume that the TS's are instantaneously created following the passage of charged particles.
The equivalent increase in local temperature
as a function of distance $b$ by a charged particle moving with velocity $v$ can be calculated after integration over all normal modes
of these energy depositions. It can be calculated directly by the average energy excitation on a molecule
\begin{eqnarray}
\Delta T (b, v) =  \frac{1}{k_B} \langle U_\gamma \rangle_\tau  (b, v).
\label{eq25rx}
\end{eqnarray}
or with a macroscopic mean value calculated by
\begin{eqnarray}
\left\langle\Delta T\right\rangle (b, v) =  \frac{1}{c_V} \frac{2\pi L}{m'(b)} \int_{0}^{b} b' db'\langle U_\gamma \rangle_\tau  (b', v).
\label{eq25r}
\end{eqnarray}
Here $c_V$ is the water-specific heat capacity at constant volume and $m'(b)$ is the mass of water embedded in a cylinder with radius $b$, and longitudinal length, $L$.
In the calculation of the stopping power, $b$ is the impact factor.
$\langle \cdots \rangle_\tau$ denotes the time averaging over a time-dependent distribution function, embedded in the time integration over the atomic excitation power-loss.
$\langle U_\gamma \rangle_\tau$ describes the energy transferred to medium as a function of distance $b$ by a charged particle moving with velocity $v$.
In MD, Eq.(\ref{eq25rx}) can be used for the initial condition and distribution of KE among the atoms within TS.
As a function of time, this initial excessive KE decays and stores randomly into PE given by the FF among all atoms including the ones surrounding TS, following the equipartition theorem in interacting many particle systems.
Whereas the usage of water heat capacity in Eq.(\ref{eq25r}) assumes such relaxation and storage of the initial thermal energy equally into KE and PE of the particles in the TS.

In a simple harmonic oscillator (SHO) model of an atom, the rate of change in the kinetic and potential energies of a valence electron, the energy stored by the charged particle can be dissipated by the rate $\gamma$ to the recoil nucleus KE. This is the dark channel of energy dissipation that is five orders of magnitude faster than the bright version of that channel (optical photon emission), according to our QM calculation. Thus, the former excitation converts to heat in the environment at a rate five orders of magnitude higher than the latter [\onlinecite{Ramin_Martin:Unpublished}].

It is a straightforward calculation to
derive analytically the following equation by
taking into account the dissipation term perturbatively (small $\gamma$ limit), using the solutions of SHO coupled with EM fields of a relativistic moving charged particle [see, for example, Ref. \onlinecite{Jackson1999:book}]
\begin{eqnarray}
\langle U_\gamma \rangle_\tau  (b, v) &=&
\frac{m}{\sqrt{2\pi}}
\frac{2}{\pi}\left(\frac{Z z e^2}{m v^2}\right)^2
\nonumber \\ &&
\int_{-\infty}^{\infty} d\omega
\frac{\omega^4}{(\omega^2_0-\omega^2)^2 + \gamma^2\omega^2} \frac{\gamma}{\gamma^4_L(\omega)}
\nonumber \\ &&
\left[
K^2_0\left(\frac{\omega b}{\gamma_L(\omega) v}\right) +
\gamma^2_L(\omega) K^2_1\left(\frac{\omega b}{\gamma_L(\omega) v}\right)
\right],
\nonumber \\
\label{eq24x}
\end{eqnarray}
where $m$ and $z$ are the atom's rest mass and the atomic number, respectively.
$Z$ and $v$ are the charge and velocity of the charged particle.
$\gamma_L(\omega) = 1/\sqrt{1-v^2/(c/n(\omega))^2}$ and
$K_0(x)$ and $K_1(x)$ are the modified Bessel functions of the first kind.
Here $n(\omega)$ is the refractive index of the medium; hence $c/n(\omega)$ represents the speed of light in a dispersive medium.
This equation is given in the CGS unit and is derived simply by integration over the SHO power loss,
$\langle U_\gamma \rangle_\tau = \int_{-\infty}^{\infty} dt P_\gamma =
m \gamma \int_{-\infty}^{\infty} dt \left(\frac{d\vec{r}}{dt}\right)^2$, using a uniform time-dependent distribution function in integration.
Here, $\vec{r}$ is the position of the atom, which gains excitation energy from the charged particle.

It is worth noting that the points are drawn from Eqs. (\ref{eq25r}) and (\ref{eq24x}), as
$\Delta T$ scales by $Z^2$, heavy ions such as $^{12}$C, $^{56}$Fe and even $^{197}$Au, near their Bragg peaks, cause an increase in temperature in a nanometer TS from 10$^4$ to 10$^5$, and even 10$^6$ K.
%Note that in the presence of energy dissipation, the time-reversal symmetry is broken and one should take into account both lateral and longitudinal components of the electric field into the calculation of power loss by the moving charge particle.

%Eq. (\ref{eq24x}) can be interpreted the optical absorption integral by converting $\gamma$ modulated with the $\omega$ dependence factor as the absorption coefficient $\alpha(\omega)$, where $\omega {\rm Im}\epsilon(\omega)/\epsilon_0 = c \alpha(\omega) n(\omega)$ and$n(\omega)=\sqrt{{\rm Re}\epsilon(\omega)/\epsilon_0}$. Here ${\rm Re}\epsilon(\omega)$ and ${\rm Im}\epsilon(\omega)$ are the real and imaginary parts of the dielectric constant.

\begin{figure}
\begin{center}
\includegraphics[width=1.0\linewidth]{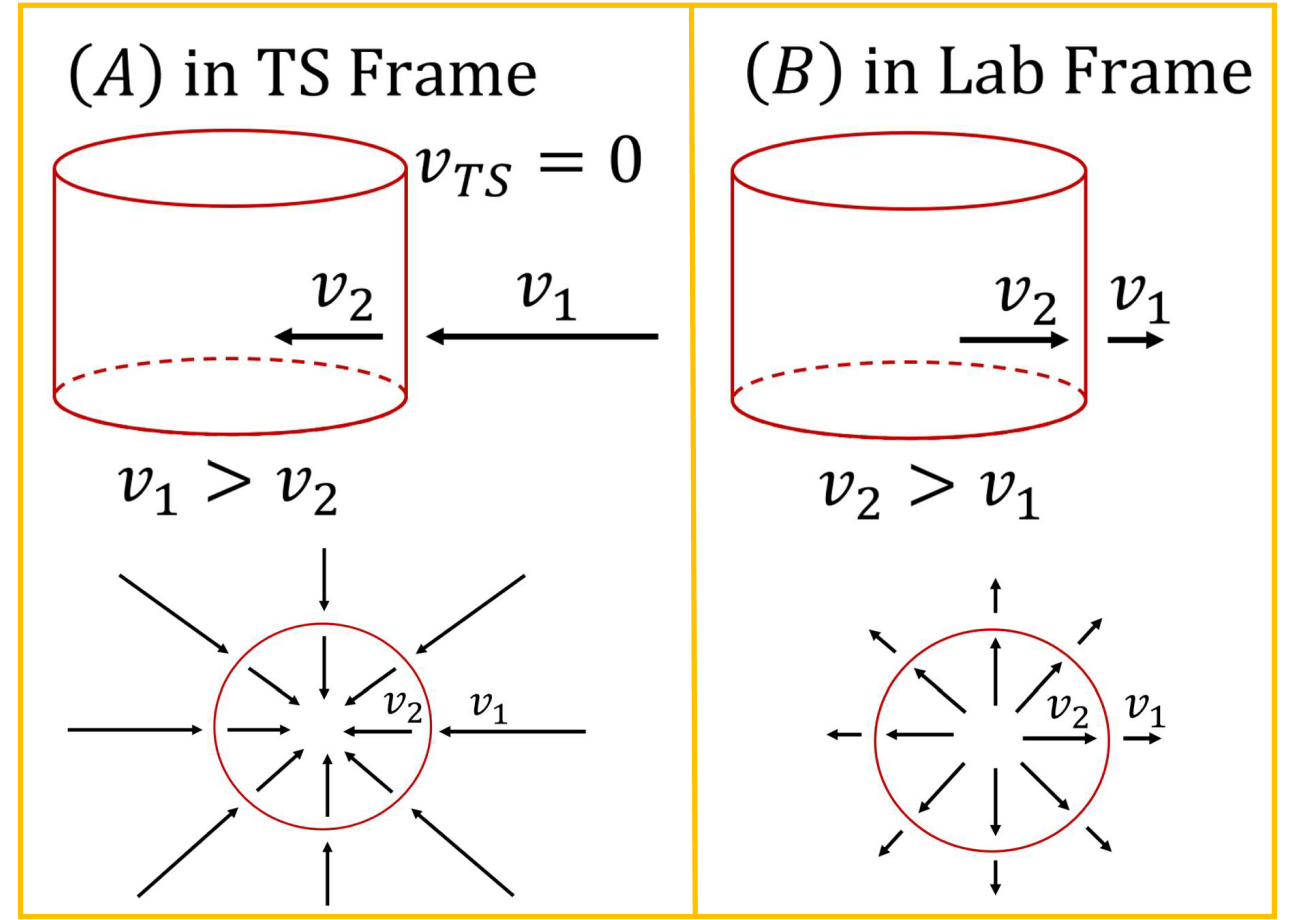}\\ %\vspace{-0.5cm}
\noindent
\caption{
Sketch of a cylindrical symmetric TS surface which is a sharp discontinuity interface separated inside and outside of a fluid denoted by labels 2 and 1 respectively.
(A) and (B) correspond to a view of an observer in a frame of references at TS (living on a flat tangent space of the cylinder) and the Lab.
$v_1$ and $v_2$ give the rate of expansion of the TS surface in (A) and (B) respectively.
The TS rest frame is from an observer's view at rest on a co-moving tangent space of a cylinder.
}
\label{Fig1}
\end{center}\vspace{-0.5cm}
\end{figure}

%%%%%%%%%%%%%%%%%%%%%%%%%%%%%%%%%%%%%%%%%%%%%%%%%%%%%%%%%%%%%

\begin{figure}
\begin{center}
\includegraphics[width=1.2\linewidth]{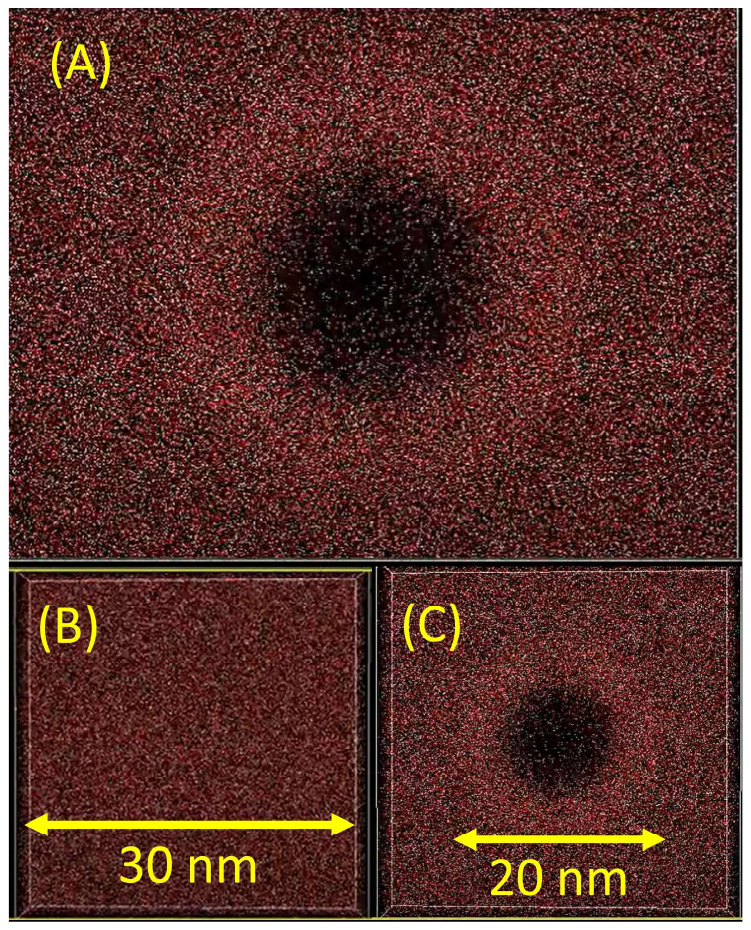}\\ \vspace{-2.5cm} % ---> Fig2
\noindent
\caption{MD simulation of half-million water molecules in the presence of a cylindrical symmetric single TS with 1 nm diameter and 0.1 fs MD time step.
The initial temperature profile that scales up proportionally to $Z^2$, corresponds to a particle with LET$=8\times 10^3$ eV/nm, e.g., $^{56}$Fe, near the Bragg peak.
The shock front propagates initially with a speed five times the speed of sound in water.
It leaves a hole behind the shock and forms a thick layer of water with a density larger than the average density of water at ambient conditions. The shock front expands to approximately 10 nm in 2 ps, with a speed equivalent of $v_s = 5,000$ m/s.
}
\label{Fig3}
\end{center}\vspace{-0.5cm}
\end{figure}

\begin{figure}
\begin{center}
\includegraphics[width=1.0\linewidth]{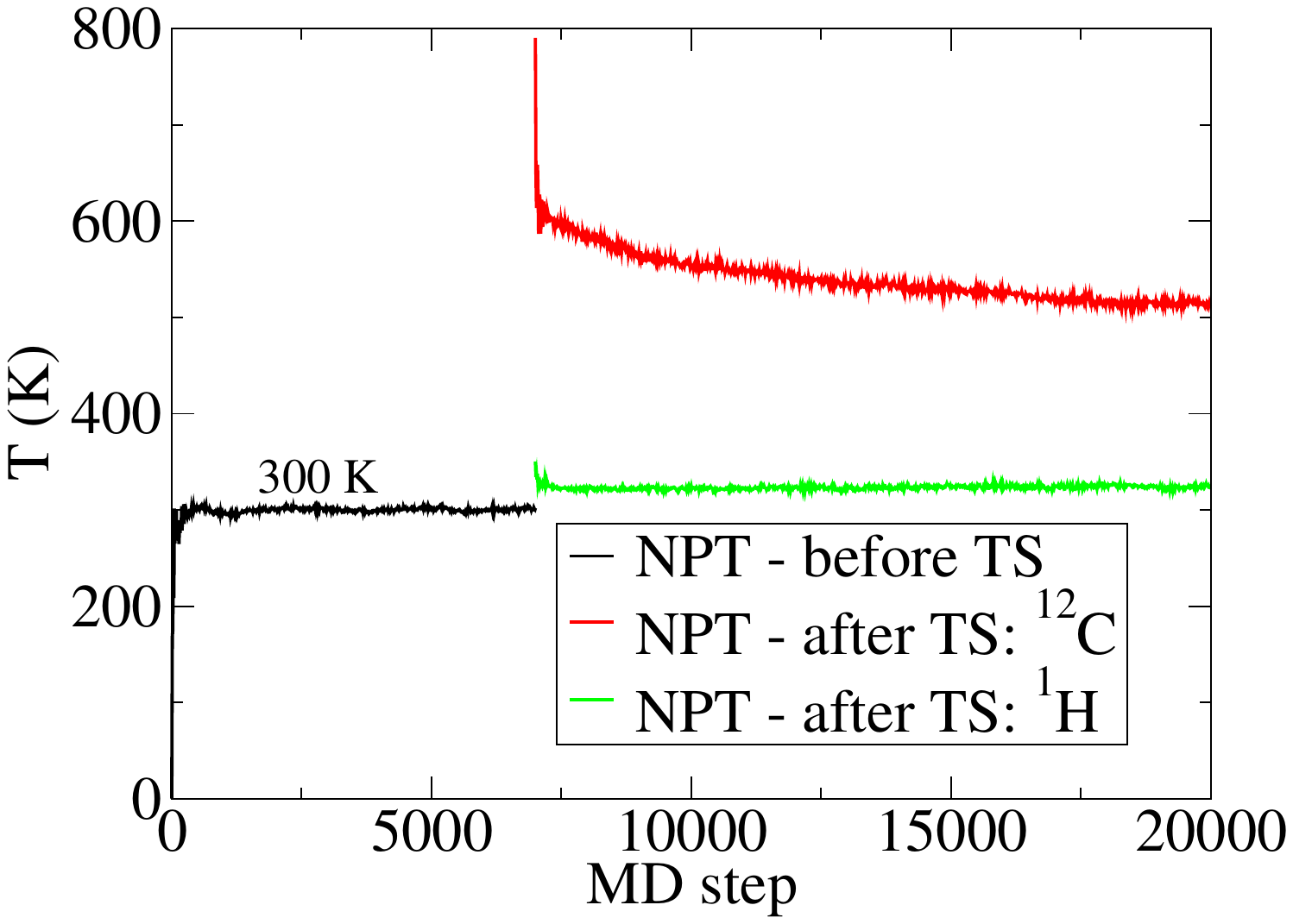}\\ %\vspace{-0.5cm} ---> Fig3
\noindent
\caption{Average temperature of the entire MD box, equivalent of mean kinetic energy per atom, as a function of MD steps for two types of charged particle,
$^{12}$C (red) and proton, $^1$H (green).
In this simulation, each step is 0.25 fs time step, thus this simulation was terminated at $t=5$ ps.
For both $^{12}$C and $^1$H we used the same computational box with PBC and 8133 water molecules and a size of $8\times 8\times 4$ nm.
The TS starts where the sharp peak is seen.
The drop from the temperature peak can be explained by the equal partitioning of the initial energy of atoms into their kinetic and potential energies.
}
\label{Fig6x}
\end{center}\vspace{-0.5cm}
\end{figure}

\subsection{Molecular dynamics}
Fig. \ref{Fig3} shows a typical state of water at a time frame after the sudden expansion due to the initial rise in the temperature of water molecules within the track of a particle, formed as a TS.
It shows the time evolution of approximately a million water molecules, a representation of a simulation for a system of over a million particles.
The interaction among the particles has been given by the parameters of ReaxFF with an MD time step of 0.1 fs using LAMMPS.

Induction of molecular thermal excitations and vibrations subsequent to the passage of charged particles in a medium results in a rapid increase of the kinetic energy and the thermal fluctuations within the nano-scale volumes, wrapping around the charged particle trajectory.
In the literature of radiation physics, this phenomenon is known as TS formation. %{\color{red}[cite old papers]}.

This, in turn, results in an abrupt variation of the local thermodynamic variables in the material embedded in a cylindrical shape, formed as an amorphous track structure.
Under high linear energy transfer, the nano-scale temperature and pressure gradients drive the molecules and chemical species in the track out of thermodynamic equilibrium relative to the rest of the background molecules, leading to a sudden expansion, burst, and generation of the chemicals in the TS.
This includes the formation of the new chemicals under extreme thermodynamic conditions just because of intensive temperature and pressure in atomic scales, in addition to ionization events and ejection of the secondary charged particles and delta-rays.

Macroscopically, the fluid dynamic equations with appropriate boundary conditions govern the time evolution of the thermodynamic variables of the system (atoms and molecules embedded in the track) coupled with the environment (rest of fluid).
For the boundary conditions, we note that the profile of energy deposition calculated by either Monte Carlo and/or amorphous track structure, predicts a gradual decrease from the center of the charged particle track.

In the present model, we however choose slightly different geometry and boundary conditions.
Without loss of generality, and for the sake of simplicity, we replace the gradual distribution of the thermodynamic variables inside the track with a surface of discontinuity due to abrupt change in pressure ($P$), number density ($\rho$),  specific volume ($V=1/\rho$), and temperature ($T$).
We consider a uniform distribution of temperature inside and outside of the track, denoted by $T_2$ and $T_1$, respectively.
Initially, at $t=0$ we consider $T_2 > T_1$ where $T_1 = 300K$ is the room temperature of the fluid.
We calculate $T_2$ by balancing the internal energy of the molecules inside the track with radius $R$ per unit length, parallel to the direction of the charged particle track, and the charged particle LET, hence $\Delta T = {\rm LET}/(A \rho_m c_V)$. Here $A$ is the cross-section area of the track. $\rho_m$ and $c_V$ are the mass density and the specific heat capacity at constant volume of unperturbed material under radiation.
We consider the ambient condition as the external condition of the material before radiation exposure.

Alternatively one may consider the balance of LET with the kinetic energy of the atoms in the cylinder using $\Delta KE = \frac{3}{2}Nk_B\Delta T$ where $\Delta KE = LET\times L$ ($L$ is the height of cylinder in MD). Note that using this formulation doubles $\Delta T$ as the energy stored in the form of PE (as presumably included in $C_V$) is initially neglected.
For illustration, let us consider the red line in Fig. \ref{Fig6x}. It shows a rise in temperature after the passage of a single track of Carbon, $^{12}$C with LET $=$ 845 eV/nm. For this calculation, we used $C_V$ to estimate the initial rise in temperature.
Considering the alternative approach, the same temperature profile may be applied to $^{12}$C with half of LET, i.e., LET = 422.5 eV/nm.

Going back to Fig. \ref{Fig3}, we used $\Delta T$ (regardless of which formulation was used for the calculation of LET corresponding to that $\Delta T$) to set up the initial conditions of the TS in the MD calculation in
a cylinder with a 1 nm diameter. The simulation shows the shock front propagates initially with a speed higher than the speed of sound in water (1450 m/s).
As seen, it leaves a hole behind the shock and forms a thick layer of
water with a density larger than the average density of water
at ambient conditions. In this calculation, the shock front expands to approximately 10 nm in 2 ps, with a speed equivalent of $v_s=5,000$
m/s.

The simulation was performed using periodic boundary conditions (PBC) in all three directions.
In PBC, there is an artifact of an interaction of the TS with its own four lateral images via compression of water molecules between the TS and the four sides of the computational box.
Thus PBC underestimates the rate of expansion of the TS's compared with a similar computation using free boundary conditions in lateral dimensions of the box and PBC in the vertical direction (parallel to the axis of the cylinder), as pointed out in Ref. [\onlinecite{deVera2019:CN}].
The speed of propagation of TS calculated in our simulations has been compared and verified with similar analytical and simulation models [\onlinecite{deVera2019:CN,deVera2018:EPJD,Surdutovich2010:PRE,Surdutovich2014:EPJD,Fraile2019:JCP}], using the same parameters as given in Fig. 4 in Ref. [\onlinecite{deVera2019:CN}].
%In response to the ionizing radiation, the medium (cells, tissues, ...) behaves like a fluid.

Regarding the comparison of the present work with early study from de Vera {\em et al.} [\onlinecite{deVera2019:CN,deVera2018:EPJD}], we remark that de Vera {\em et al.} have reported usage of both the standard CHARMM-FF and a modification to reactive CHARMM-FF in their publications for MD simulations of thermal spikes or shock waves in water in the presence of DNA. The standard CHARMM-FF is not appropriate for simulating chemical reactions. The CHARMM-FF potential (and AMBER-FF, GROMACS-FF that have been used vastly for the simulation of protein folding) is a series of simple harmonic oscillators without a transition point needed for breaking and forming bonds. Therefore the FF as such can not simulate the formation of radicals and other chemical species, in addition, it lacks simulating DNA damage (either single-strand or double-strand breaks). That is the main rationale for using ReaxFF (trained by a large series of quantum mechanical calculations) in this work and comparing with reactive CHARMM-FF reported by de Vera {\em et al.} [\onlinecite{deVera2018:EPJD}] for the TS effects on the hydrodynamics of water.
%Indeed, a series of FFs in the class of CHARMM-FF such as AMBER-FF, GROMACS-FF have been used vastly for the simulation of protein folding. The problems as such do not require breaking/forming covalent bonds, in contrast to our problem where we tend to simulate the formation of reactive oxygen species (ROS) and DNA single/double strand damage or protein/lipid breakage as a result of passage of charged particles or ionizing radiation. Besides, we expect the water molecules in the de Vera et al. to be TIP3P or a similar model which is a rigid model of water molecules, incapable of disintegration to OH radicals and other ROS's. Also, the heat capacity and other thermodynamic variables of TIP3P would deviate from the realistic model of water. In our work, we correctly calculated the thermodynamic variables for water using ReaxFF, in agreement with the experimental data, reported in NIST database [\onlinecite{NIST:HeatCapacity}].

%%%%%%%%%%%%%%%%%%%%%%%%%%%%%%%%%%%%%%%%%%%%%%%%%%
\subsection{Interface dynamics}
Following the above conventions, we denote $P_1, \rho_1, T_1$, the thermodynamic variables at a point outside of the track interface, infinitesimally close to the surface of discontinuity as shown in Fig. \ref{Fig1}.
Similarly $P_2, \rho_2, T_2$ denote the same set of thermodynamic variables, at a similar point but on the opposite side of the interface, inside of TS with an infinitesimal distance from the surface of discontinuity. Note $V_1 = 1/\rho_1$ and $V_2 = 1/\rho_2$.
We match these two sets of variables using conventional fluid dynamic equations [\onlinecite{Landau_Lifshitz1987:Hydro_Book}].

%We now turn to match the thermodynamic variables across the thermal spike interface using the standard fluid dynamic equations of motion [Landau-Lifshitz fluid mechanics].
In the frame of reference of the TS interface (and moving laboratory frame of reference with velocity in opposite direction) the conservation of mass, energy, and momentum flux is given by
\begin{eqnarray}
{\rm mass:} ~~~~~~ \rho_1 v_1 = \rho_2 v_2 \equiv J,
\label{eq1}
\end{eqnarray}
\begin{eqnarray}
{\rm energy:} ~~~~~~ H_1 + \frac{1}{2} v^2_1 = H_2 + \frac{1}{2} v^2_2,
\label{eq2}
\end{eqnarray}
\begin{eqnarray}
{\rm momentum:} ~~~~~~ P_1 + \rho_1 v^2_1 = P_2 + \rho_2 v^2_2.
\label{eq3}
\end{eqnarray}
Here $v_1$ and $v_2$ are the components of the velocities normal to the thermal-spike cylinder, and $J$ is the conserved fluid current number density.
$H = \tilde{\gamma}/(\tilde{\gamma}-1) P/\rho$ is the enthalpy, independent of the equation of the state of the fluid.
Here $\tilde{\gamma}=C_P/C_V$.
We emphasize that this expression for $H$ is not based on the ideal gas equation of state.

The system of coupled equations for the conservation laws
%plus the equation of state for the fluid, e.g., the perfect gas $PV = RT$ ($V = 1/\rho$),
can be solved together by straightforward algebraic manipulations in terms of the Mach number, $M = v_1 / c_s$ and $\tilde{\gamma}$.
%Here $v_s = v_1$ is the SW velocity, right at the SW discontinuity, defined on the surface of TSs from outside, in the unperturbed side of the fluid.  $c_s = 1/\sqrt{\rho_1\beta_{T1}}$ is the speed of sound in front of SW, in the side of the fluid denoted by index 1, outside of the track where the fluid is unperturbed (see Fig. \ref{Fig1}).
As mentioned above, the side denoted by 1, is in front of the TS interface, on the unperturbed side of the fluid.
The side denoted by 2, behind the TS interface and inside the TS volume, may undergo a phase transition and form a hot and dilute gas with different chemical compositions.

%Note that $v_2$ has a profile from the center of TS to the surface of SW. $v_1$ can be zero everywhere but right on the surface of TS. As a function of time, the TS expansion rate described by $v_1$ at the surface can decay to the speed of sound. From that moment, the surface of TS propagates with the speed of sound, $c_s$.

%Neglecting flow of the fluid in front of SW, assuming $v_1=0$ (except at the surface of SW) in the Lab frame of reference,

Using the continuity equations, and calling $H = \tilde{\gamma}/(\tilde{\gamma} - 1) (P/\rho)$ with
$\tilde{\gamma}=C_P/C_V$, assuming $\tilde{\gamma}_1 = \tilde{\gamma}_2$
one can relate the thermodynamic variables right behind the TS surface, $(\rho_2, P_2)$, to the same thermodynamics variables right in front of the TS surface, $(\rho_1, P_1)$, two points adjacent to each other but in opposite sides of the TS surface.
\begin{eqnarray}
\frac{\rho_2}{\rho_1} = \frac{v_1}{v_2} = \frac{(\tilde{\gamma}+1) \tilde{M}^2}{(\tilde{\gamma} +1) + (\tilde{\gamma} - 1)(\tilde{M}^2 - 1)} \approx \frac{\tilde{\gamma} + 1}{\tilde{\gamma} - 1},
\label{eq4x}
\end{eqnarray}
\begin{eqnarray}
\frac{P_2}{P_1} = \frac{(\tilde{\gamma} +1) + 2\tilde{\gamma}(\tilde{M}^2 - 1)}{\tilde{\gamma} + 1} \approx \frac{2\tilde{\gamma}}{\tilde{\gamma} + 1} \tilde{M}^2.
\label{eq5x}
\end{eqnarray}
Here $\tilde{M} = v_1/(\lambda_1 c_s) = v_1/\tilde{c}_s$ where $\lambda_1 = \sqrt{P_1 \beta_{T,1}}$ is the speed of sound correction factor due to incompressibility of the fluids, $\beta_{T,1} = V^{-1}(\partial V/\partial P)_T$ is the fluid compressibility at point 1,
and $\tilde{c}_s = \lambda_1 c_s = \sqrt{\frac{\tilde{\gamma}_1  P_1}{\rho_1}}$ that is the equation for the speed of sound in the ideal gases, but because of different values of the thermodynamic parameters using the incompressible fluids at side 1 in this equation, $\tilde{c}_s$ doesn't match with the empirical values of speed of sound in ideal gases.
Thus %$\tilde{M} = \frac{1}{\lambda_1} M$
\begin{eqnarray}
\tilde{M} = \frac{1}{\lambda_1} M
\label{eq5xx}
\end{eqnarray}
does not represent Mach number unless the fluid inside 1 would be an ideal gas.

To relate the temperatures $T_2$ to $T_1$, we must know the equation of state of the fluid.
Here, to proceed with the derivation of the analytical solutions, let us assume the perfect gas equation of the state, $PV = RT$ ($V = 1/\rho$). This simplification of approximating water to an ideal gas (which is not correct) allows us to imply
$H = \tilde{\gamma}/(\tilde{\gamma} - 1) (P/\rho) = \tilde{\gamma}/(\tilde{\gamma} - 1) k_B T$.
\begin{eqnarray}
\frac{T_2}{T_1} &=& \frac{[(\tilde{\gamma}+1)+2\tilde{\gamma}(\tilde{M}^2 - 1)][(\tilde{\gamma} +1) + (\tilde{\gamma} - 1)(\tilde{M}^2 - 1)]}{(\tilde{\gamma}+1)^2\tilde{M}^2}
\nonumber \\ &&
\approx \frac{2\tilde{\gamma}(\tilde{\gamma}-1)}{(\tilde{\gamma}+1)^2} \tilde{M}^2.
\label{eq6x}
\end{eqnarray}
Here $\tilde{M} = M$ as we have used the ideal gas equation of state to derive Eq.(\ref{eq6x}).
We recall that $R=N_A k_B$ therefore $\rho^{-1} = V/N_A$ is the gas-specific volume (here $V$ is the total volume of the gas).
Finally, by substituting the density with specific volume, $V$, where $\rho = 1/V$ in Eq. (\ref{eq4x}), hence, $\rho_2 / \rho_1 = V_1 / V_2$, we can check the consistency of Eqs. (\ref{eq4x}-\ref{eq6x}) after multiplying Eqs. (\ref{eq4x}) to (\ref{eq5x}) which leads to Eq. (\ref{eq6x}), hence
\begin{eqnarray}
\frac{T_2}{T_1} = \frac{P_2}{P_1} \frac{V_2}{V_1}.
\label{eq6_1}
\end{eqnarray}

%Note that as $\tilde{\gamma} \neq \gamma_S$, the SW equations are not isentropic.

From the RHS of Eqs. (\ref{eq4x}), (\ref{eq5x}), and (\ref{eq6x}), a supersonic motion of particles localized on the TS interface is predicted in the limit of high LET, where $\tilde{M} >> 1$.
In this limit, the pressure inside the thermal-spike volume grows quadratically by $\tilde{M}$.
In addition, $P_2 > P_1$, $\rho_2 > \rho_1$, $T_2 > T_1$, hence $v_2 < v_1$ in TS frame of reference.
See Fig. \ref{Fig1}.

In this limit, the rate of expansion of a cylindrical TS assuming a perfect gas equation of the state of the fluid can be calculated analytically.
To this end, from Eq. (\ref{eq5x}) we find $1/T_1 \propto \tilde{M}^2 = v_1^2/\tilde{c}_s^2$ (here $\tilde{c}_s=c_s$ thus $\tilde{M} = M$).
As $v_1 = \dot{R}$, where $R$ is the radius of the cylinder that encompasses the TS surface, we obtain $1/T_1 \propto \dot{R}^2$.
For ideal gases, $T_1 \propto V_1$ where $V_1$ is the volume embedded the cylinder, $V_1 \propto R_1^2$, thus $1/T_1 \propto 1/V_1 \propto 1/R^2$.
Combining these we obtain $\dot{R}^2 \propto 1/R^2$ and $R dR \propto dt$ which lead to $R^2 \propto t$ or $R \propto \sqrt{t}$.
These are solutions of strong explosion in cylindrical symmetry, calculated using scaling arguments and dimensional analysis [\onlinecite{Surdutovich2010:PRE}].
The same calculation, as described here, predicts $R \propto t^{2/5}$ for a spherical surface as $V_1 \propto R^3$ originally calculated by Sedov and von Neumann [\onlinecite{Sedov1946:PMM,vonNeumann1947:Book,Zeldovich1996:Book}].

As a function of time, the dissipation mechanisms of supersonic motion prevail and temperature-pressure gradients damp out rapidly until $M$ drops to unity asymptotically where $\rho_2=\rho_1$, $v_2=v_1$, $P_2=P_1$, and $T_2=T_1$.
From that time, the interface that has evolved into a thin volume propagates like an amplitude of perturbed wave in pressure and temperature with the speed of sound without transporting the molecules collectively.
We will evaluate the speed of sound of such a system by performing MD in the next section.

For an imperfect gas or a liquid with $1/(P\beta_T) >> 1$, the condition $\tilde{M} = 1$ where $v_1 = \tilde{c}_s$ in Eqs. (\ref{eq4x}) and (\ref{eq5x}) leads to the following thermodynamic conditions $\rho_2=\rho_1$, $v_2=v_1$, and $P_2=P_1$, respectively.
For a perfect gas, these conditions can be reached at a Mach number equal to 1, that is, as soon as $v_1$ reaches the speed of sound in the gas.
But for incompressible fluids, such as water, the TS dissipates to a sound wave at the wavefront speed and continues to expand even with a speed below the speed of sound, as in $c_s > \tilde{c}_s$.

\subsection{Thermodynamic variables}
The dynamical formulation of TS expansion in a continuous fluid, presented in the preceding section assumes the surface of TS's, a sharp moving discontinuity.
However, at the microscopic level, the TS interface consists of particles with a finite size thickness. %, like a domain wall.
Away from the interface, inside TS, we expect the formation of a diluted and/or gas phase cavity with a structure sketched in Fig. \ref{Fig3} where the density drops below $\rho_1$ with temperature and pressure above $T_1$ and $P_1$.

In general, the time evolution of the equation of the state of the fluid is unknown.
Many authors simplify the model by considering the fluid as a perfect gas.
To avoid such simplifications and to make the model as realistic as possible, we perform an MD simulation on a large number of molecules to be able to handle uncertainties in the equation of the state of the water in the presence of TSs, presumably induced by the passage of charged particles.
A representative example of the simulation is shown in Fig. \ref{Fig3}.
To capture the dynamical expansion of a typical TS in real-time, we have uploaded samples of our simulations and videos, available on YouTube: [\onlinecite{Abolfath2023:YouTube,Abolfath2023_2:YouTube,YouTube_two_SWs}].

In MD, as a function of time and space, we calculate the local thermodynamic variables such as mass density ($\rho$), temperature ($T$), pressure ($P$), specific heat capacities, $c_P = C_P/m$, $c_V = C_V/m$, isentropic index $\gamma_S$ where the index $S$ refers to entropy,
isothermal compressibility $\beta_T$ and its inverse, the bulk modules, and finally the speed of sound $c_s$ in the fluid.
We denote $\lambda^2 = P\beta_T$, a correction factor in the transformation between ideal gas and real fluids, needed for TS boundary conditions in the equations for conservation of mass, momentum, and energy, as described above.

We perform this calculation in two ways: (1) ensemble averaging over the entire system and taking all atoms into account and (2) partitioning the computational box into small volume elements, and voxels, and using them as statistical bins.
These volumes are equivalent to grand canonical ensembles where the exchange of energy and particles occurs in their boundaries.

Numerically it is straightforward to calculate the average mass ($m$) and kinetic energy ($K.E.$) in each volumetric element, at each time slice, to obtain $\rho(\vec{r}, t) = \langle m(\vec{r}, t) \rangle/V$, and
$\langle K.E.(\vec{r}, t) \rangle = 3k_B T(\vec{r}, t)/2$.

In MD the local pressure ($P$) and its deviation from the ideal gas (system of non-interacting particles) is calculated using the Virial equation [\onlinecite{HuangSatPhys,Lion2012:JPC}]
\begin{eqnarray}
P(\vec{r}, t) &=& \rho(\vec{r}, t) k_B T(\vec{r}, t) + \frac{1}{3V} \left\langle \sum_{i<j}^{N} \vec{f}_{ij}(\vec{r}, t) \cdot \vec{r}_{ij}(\vec{r}, t)\right\rangle.
\nonumber \\ &&
\label{eq00}
\end{eqnarray}
Here $\vec{f}_{ij}$ is a pair-wise force acting between particles $i$th and $j$th with separation $\vec{r}_{ij} = \vec{r}_{i} - \vec{r}_{j}$.
The state of an ideal gas with negligible $\vec{f}_{ij}$ is given by the first term in Eq. (\ref{eq00}).
Note that $\vec{r}$ refers to the location of volumetric elements.

It would be useful to present a review of the relations between thermodynamic variables, relevant to our simulations.
The enthalpy and the internal energy of the system are given by $H = E + PV$ and U = E = K. E. + P. E., thus by definition,
%\begin{eqnarray}
$C_V = \left(\frac{\partial U}{\partial T}\right)_V =
\left(\frac{\partial U}{\partial T}\right)_P + P \left(\frac{\partial V}{\partial T}\right)_p$,
%\end{eqnarray}
and
%\begin{eqnarray}
$C_P = \left(\frac{\partial H}{\partial T}\right)_P =
\left(\frac{\partial U}{\partial T}\right)_P + P \left(\frac{\partial V}{\partial T}\right)_p$.
%\end{eqnarray}
At constant $P$, the internal energy is only a function of $T$ and $V$, thus $U = U(T, V)$ then $dU = \left(\frac{\partial U}{\partial T}\right)_V dT + \left(\frac{\partial U}{\partial V}\right)_T dV$.
Applying a chain rule of partial differentiation we obtain
%\begin{eqnarray}
$\left(\frac{\partial U}{\partial T}\right)_P =
\left(\frac{\partial U}{\partial T}\right)_V +
\left(\frac{\partial U}{\partial V}\right)_T \left(\frac{\partial V}{\partial T}\right)_P$.
%\end{eqnarray}
Hence
%\begin{eqnarray}
$C_P =
\left(\frac{\partial U}{\partial T}\right)_P + P \left(\frac{\partial V}{\partial T}\right)_p
%\nonumber \\ &=&
=\left(\frac{\partial U}{\partial T}\right)_V +
\left(\frac{\partial U}{\partial V}\right)_T \left(\frac{\partial V}{\partial T}\right)_P
+ P \left(\frac{\partial V}{\partial T}\right)_P
%\nonumber \\ &=&
= C_V +
\left[P + \left(\frac{\partial U}{\partial V}\right)_T\right]\left(\frac{\partial V}{\partial T}\right)_P$.
%\end{eqnarray}
In non-interacting ideal gas, $U=KE$ and $PE=0$. Thus the mean value of $U$ is only a linear function of temperature, $T$, and it is independent of volume $V$, hence $\left(\frac{\partial U}{\partial V}\right)_T = 0$. Applying the equation of the state for ideal fluids, $PV = nRT$ gives
%\begin{eqnarray}
$C_P - C_V =
P\left(\frac{\partial V}{\partial T}\right)_P = nR$
%\end{eqnarray}
with $n = N/N_A$. $N$ and $N_A$ are the number of particles and Avagadro's number respectively.
Defining per molar heat capacities, $C_P/n$ and $C_V/n$, gives $C_P - C_V = R$.

For a system of interacting particles, one can show (using thermodynamic's free energies and Maxwell relations)
%\begin{eqnarray}
$C_P = C_V + V T \frac{\alpha^2_P}{\beta_T}$,
%\end{eqnarray}
where $\alpha_P = \frac{1}{V} \left(\frac{\partial V}{\partial T}\right)_P$ and
$\beta_T = - \frac{1}{V} \left(\frac{\partial V}{\partial P}\right)_T$
are the isobaric expansion coefficient and isothermal compressibility respectively.
For incompressible fluids  (including water), $\alpha$ and $\beta$ are negligible thus
$C_P/C_V \approx 1$ as we found in this study.

The isentropic (adiabatic) index, was calculated using  $\gamma_S = \left(\partial H/\partial U\right)_S = (C_P/C_V)/(P \beta_T)$.
For ideal gas $P \beta_T=1$ hence $\gamma_S = \tilde{\gamma} = C_P/C_V$ where $\tilde{\gamma}$ used to describe the TS wavefront in the preceding section.
For non-ideal fluids, $\lambda^2 = P \beta_T$, calculated in MD, is the correction factor that must be applied to $\gamma_S$.

\begin{figure}
\begin{center}
\includegraphics[width=1.0\linewidth]{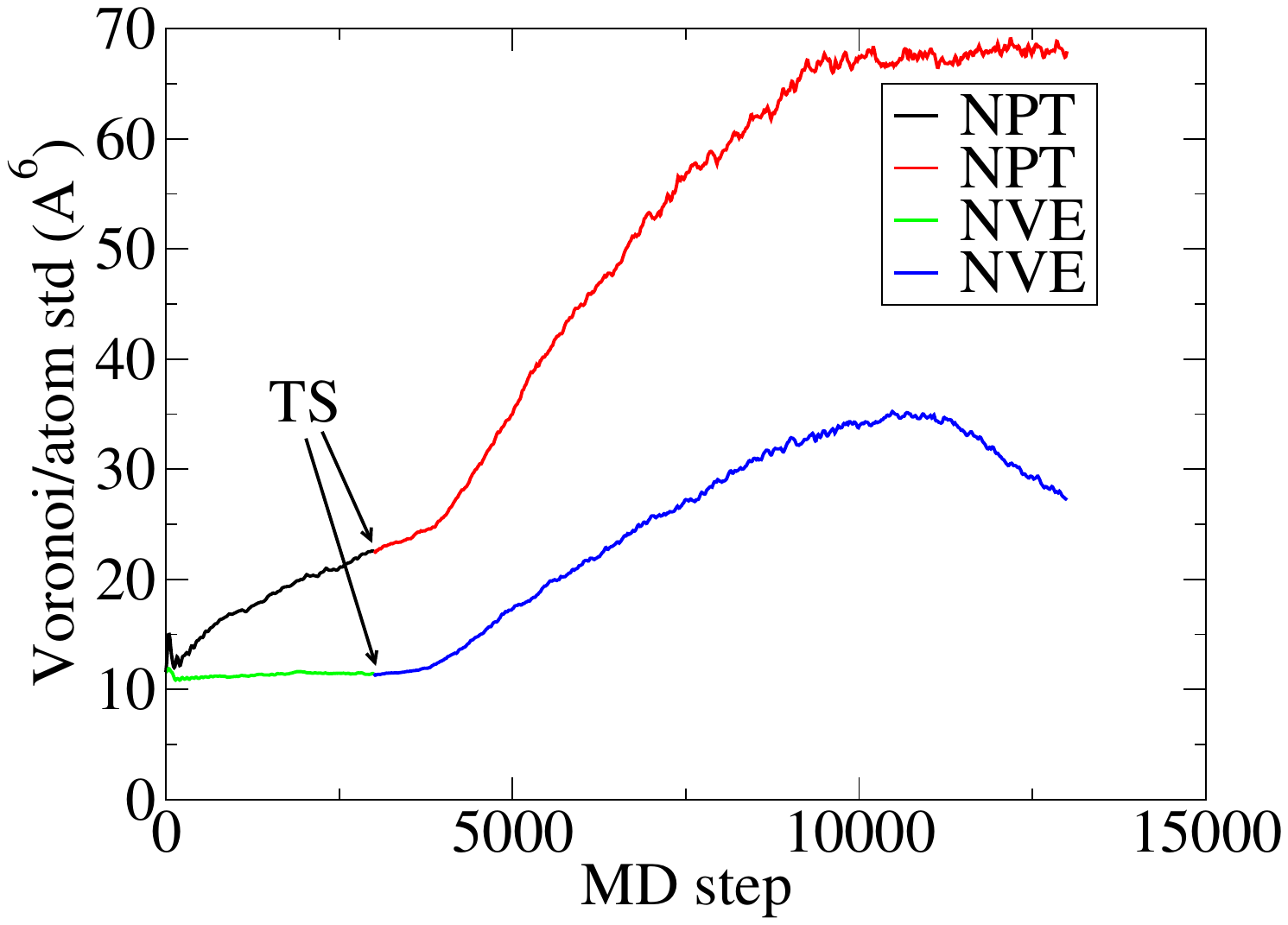}\\ %\vspace{-0.5cm} ---> Fig4
\noindent
\caption{STD of volumes per atom computed by Voronoi-method in LAMMPS.
}
\label{Fig7x}
\end{center}\vspace{-0.5cm}
\end{figure}

To calculate the thermodynamic variables listed above, we use well-established relations that correlate the fluctuations of the variables calculated in a finite MD system and their thermodynamic counterparts.
In NVE microcanonical and/or NPT ensembles.
For more details, see page 65 [\onlinecite{Chandler:book}], eq.(2.70) in [\onlinecite{AllenTildesley:book}], and [\onlinecite{LandauLifshitz:SatPhysbook}] (page. 333 on a general concept of Fluctuations in thermodynamics) where
\begin{eqnarray}
C_V = \frac{\langle E^2\rangle - \langle E\rangle^2}{k_B T^2}
\end{eqnarray}
where $E$ is the atomistic total energy, the sum of the particles' kinetic and potential energies, $E = KE + PE$.
Similarly to calculate $C_P$ and $\beta_T$  (Eqs. 2.85 and 2.86 in [\onlinecite{AllenTildesley:book}]) in NPT ensemble we use
\begin{eqnarray}
C_P &=& \frac{\langle H^2 \rangle - \langle H \rangle^2}{k_B T^2}
\nonumber \\ &=&
\frac{\langle (E + PV)^2 \rangle - \langle E+PV \rangle^2}{k_B T^2}
\nonumber \\ &=&
\frac{\langle E^2 \rangle - \langle E \rangle^2}{k_B T^2}
+
P^2 \frac{\langle V^2 \rangle - \langle V \rangle^2}{k_B T^2}
\nonumber \\ &+&
2P \frac{\langle E V \rangle - \langle E\rangle\langle V\rangle}{k_B T^2}
\end{eqnarray}
and
\begin{eqnarray}
\beta_T = \frac{1}{k_BT}\frac{\langle V^2 \rangle - \langle V \rangle^2}{V}
\end{eqnarray}
Note that the the total volume in the NTP ensemble fluctuates whereas in NVE is constant.
In MD, we calculate Voronoi as a metric for the inter-atomic occupation space and the specific volume per atom, $V_{\rm vor}$, a variable as a function of time in both NPT and NVE ensembles.
It satisfies a geometrical sum rule, $V = \sum_{i=1}^N V_{{\rm vor}, i}$,
where $N$ is the number of atoms in the computational box. Hence, $\delta V^2 = \langle V^2 \rangle - \langle V \rangle^2$ where
$\langle V^2 \rangle = \sum_{i=1}^{N} V_{\rm vor, i}^2 / N$ and $\langle V \rangle = \sum_{i=1}^{N} V_{\rm vor, i} / N$.
The results of this calculation for two ensembles of NPT and NVE before and after the introduction of a single TS are shown in Fig. \ref{Fig7x}.
It is interesting to note that the transition before and after the addition of TS is continuous in the atomic-specific volumes.
We also observe that fluctuations in $V$ are substantially smaller in NVE compared to the NPT ensemble, as expected.
%Thus in NVE ensemble, the last two terms in the above equation vanishes and the resulting heat capacity calculated in LAMMPS converges to $C_V$.

\begin{figure}
\begin{center}
\includegraphics[width=1.0\linewidth]{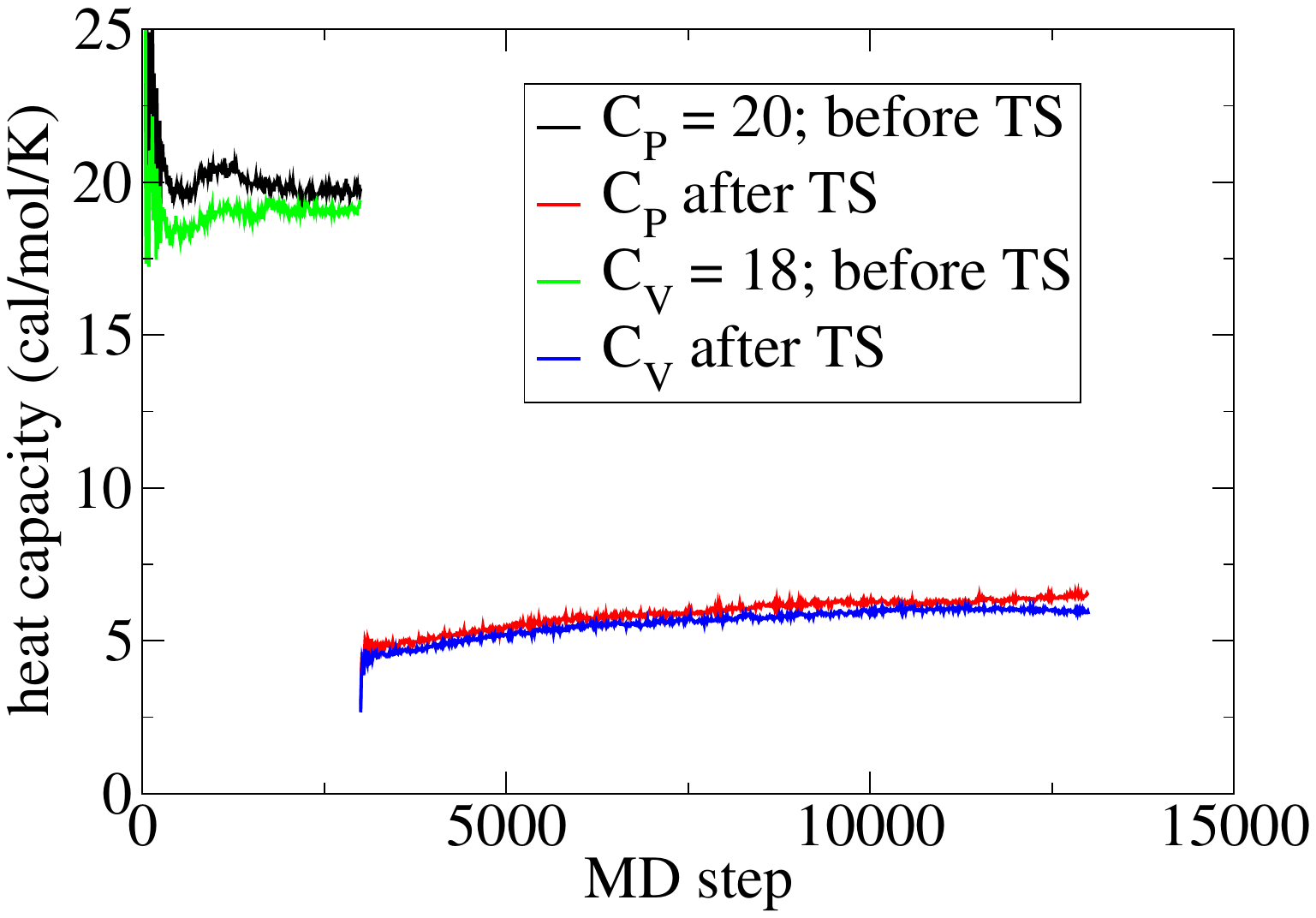}\\ %\vspace{-0.5cm} ---> Fig5
\noindent
\caption{Heat capacities calculated for the entire computational box at constant pressure and volume using NPT and NVE ensembles.
A sharp drop in $C_P$ and $C_V$ occurs at the onset of TS formation.
This is an indication of a first-order phase transition and the emergence of a dilute gas of water molecules mixed with other chemical species inside the nano-cavity.
The predicted values of heat capacities at room temperature (300 K) show a good match with published NIST data.
% https://webbook.nist.gov/cgi/cbook.cgi?ID=C7732185&Units=CAL&Type=JANAFL&Table=on
}
\label{Fig8x}
\end{center}\vspace{-0.5cm}
\end{figure}

As shown in Fig. \ref{Fig8x}, our MD calculation of the specific heat capacities of the bulk water with ReaxFF at ambient conditions, and before creation of the TS's, converges to $C_V = 18~{\rm Cal/mol/K}$ and $C_P = 20~{\rm Cal/mol/K}$.
The initial fluctuations at the beginning of the simulations are the steps in MD before reaching the system to the thermal equilibrium at room temperature.

After the creation of TS's, a diluted state of water, mixed with chemical species that are the products of dissociation of water molecules due to high-temperature fluctuations, are formed, and an abrupt drop in heat capacities is seen as an indication of a first-order phase transition.

These values are in agreement with the reported experimental data from NIST [\onlinecite{NIST:HeatCapacity}].
Thus we obtain $\tilde{\gamma}=C_P/C_V=1.11$, hence the condition for the strong explosion [\onlinecite{Landau_Lifshitz1987:Hydro_Book}] can be obtained if the ratio of internal to external pressure in TS's is satisfied $P_2/P_1 > \rho_2/\rho_1 = (\tilde{\gamma} + 1)/(\tilde{\gamma} - 1) = 19$. See, e.g., Eqs. (\ref{eq4x}) and (\ref{eq5x}) in the limit of $\tilde{M} >> 1$).
Our calculated value for $\tilde{\gamma}=1.11$ is close to the value given by de Vera {\em et al.} [\onlinecite{deVera2019:CN}],  $\tilde{\gamma}=1.22$.
However, the predicted strong explosion criteria using this value reduces the lower bound of the pressure ratio to 10.

To compare the numerical values with the prediction of adiabatic expansion we use $c_s = 1450$ m/s in water and the experimental value of the specific heat that predicts $\gamma_S \approx 2.9$ which is three times larger than reported $C_P/C_V \approx 1.11$.
With these considerations, one may expect deviations from Sedov's solutions [\onlinecite{Sedov1946:PMM,Surdutovich2010:PRE}] obtained analytically through ideal gas assumptions.
%This indicates an overestimation of adiabatic expansion of the shock waves under the self-similar solutions such as Sedov's solutions derived elegantly in [PRE 2010].

\begin{figure}
\begin{center}
\includegraphics[width=1.0\linewidth]{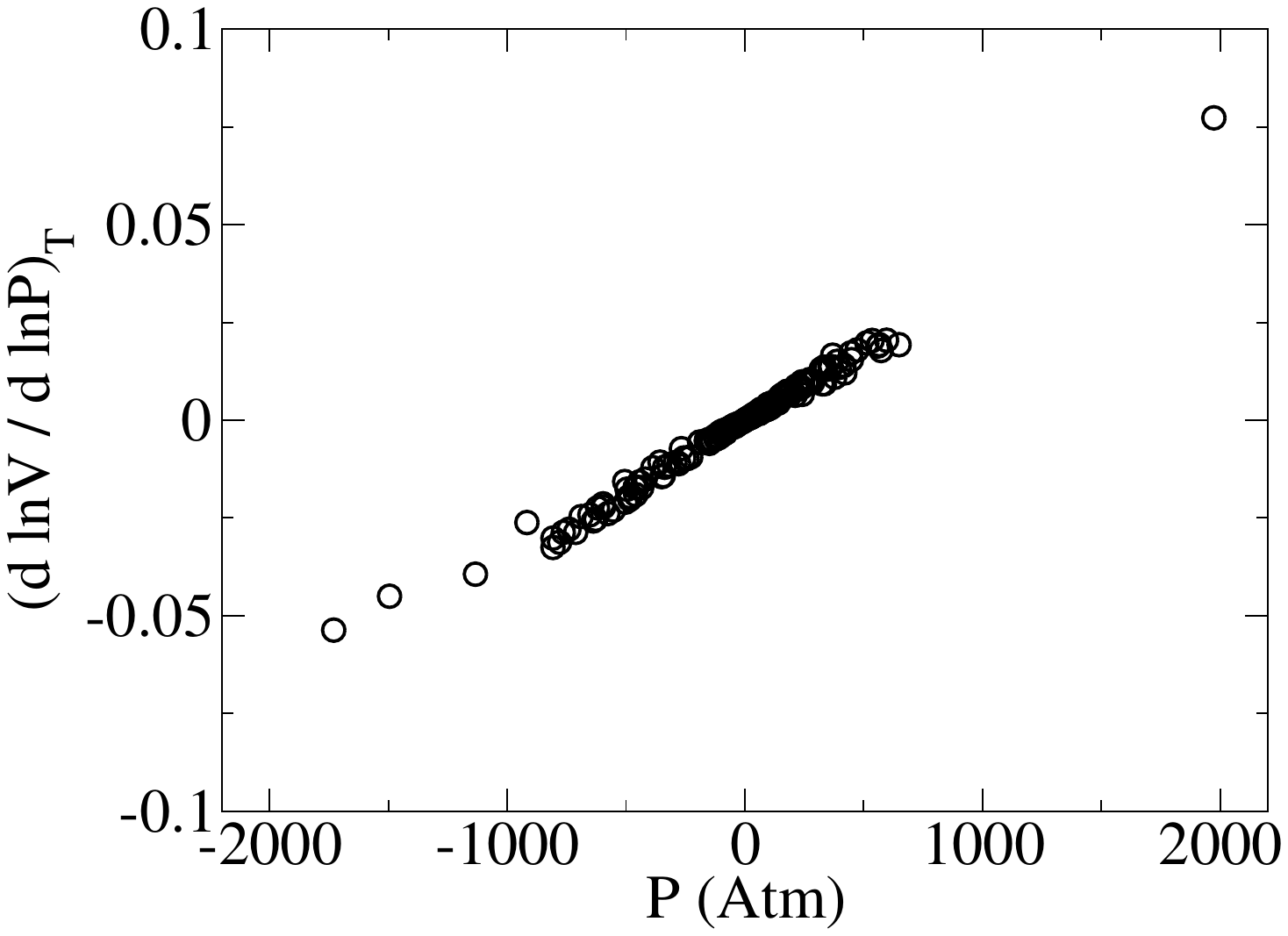}\\ %\vspace{-0.5cm} ---> Fig6
\includegraphics[width=1.0\linewidth]{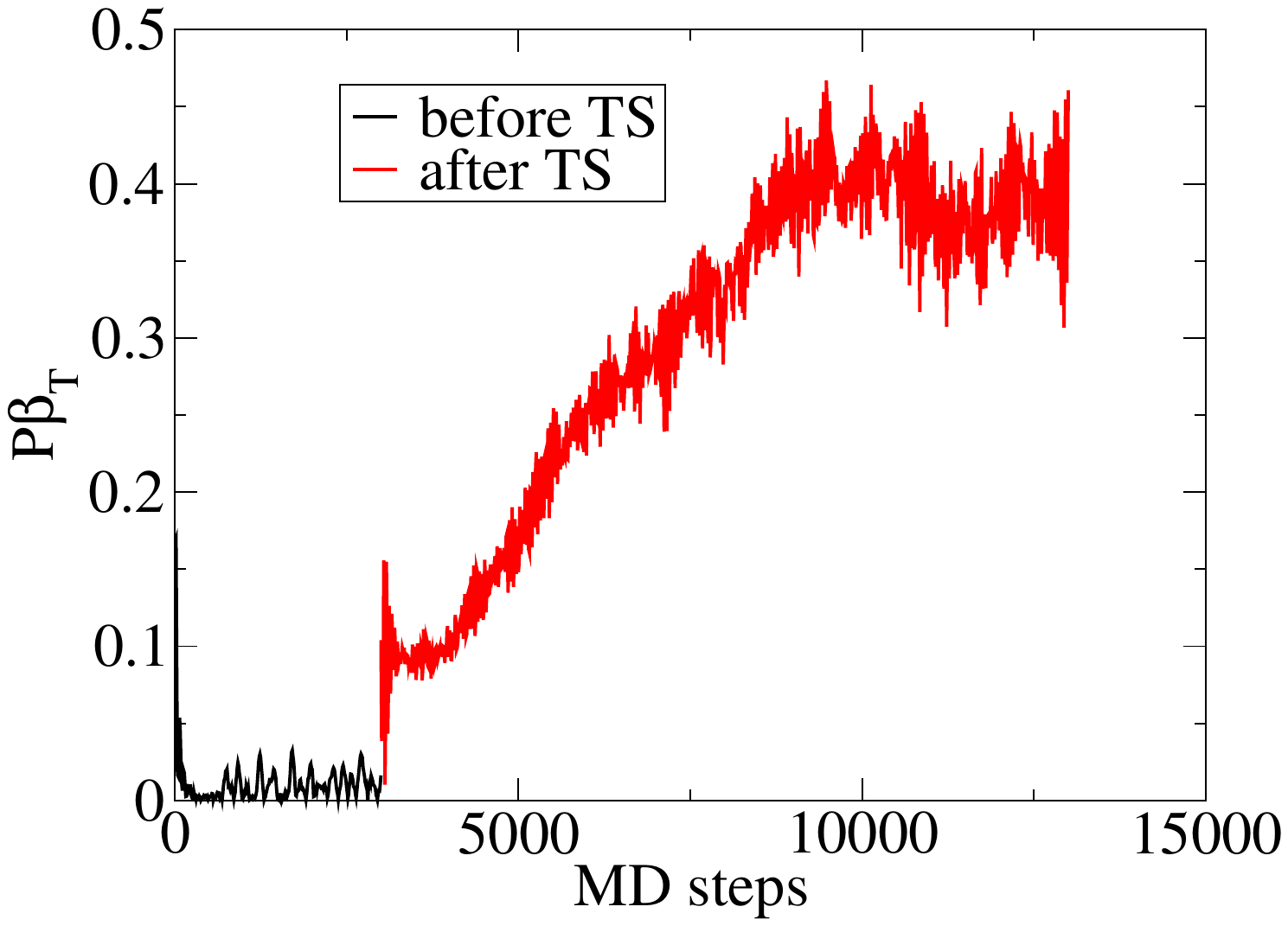}\\ %\vspace{-0.5cm}      ---> Fig7
\noindent
\caption{Top: Calculation of $P\beta_T = (P/V)(\partial V/\partial P)_T$ as a function of pressure for normal water is shown.
As $\gamma_S = (C_P/C_V)/(P\beta_T)$, strong deviation from ideal gas due to incompressibility of liquid water is expected at low pressures.
Bottom: $P\beta_T$ vs. MD steps. It seems the system of TS fluctuates around $P\beta_T = 0.4$, lower than the corresponding value for ideal fluids ($P\beta_T = 1$ for ideal gas).
}
\label{Fig9x}
\end{center}\vspace{-0.5cm}
\end{figure}

Fig. \ref{Fig9x} shows the result of our MD calculation for the isothermal incompressibility in the NPT ensemble as a function of pressure in the water and MD steps (time).
Before TS generation, $P\beta_T = (P/V)(\partial V/\partial P)_T$ approaches a small value due to the incompressibility of water. The small fluctuations are seen because of the local pressure controlling the Nosé–Hoover thermostat and barostat to equilibrate the system to the requested $T$ and/or $P$.
After TS's though we initially observe strong fluctuations in $P\beta_T$ which quickly saturates to an asymptotic value of $P\beta_T\approx 0.4$. This is because of a diluted gas of water inside TS, mixed with a condensed state of water on the shell and outside of TS.
We note that for an ideal gas $P\beta_T=1$.

\begin{figure}
\begin{center}
\includegraphics[width=1.0\linewidth]{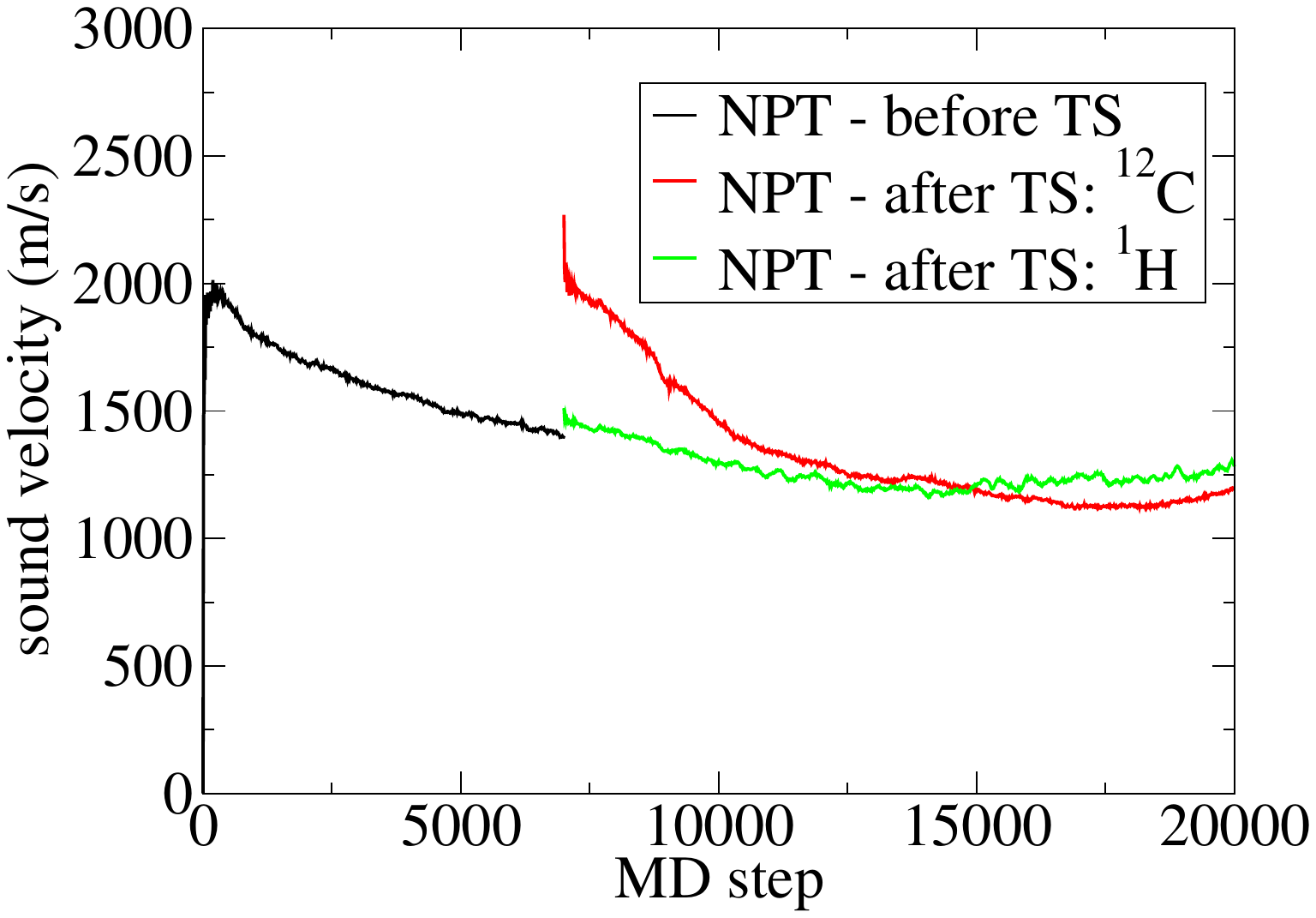}\\ %\vspace{-0.5cm} ---> Fig8
\noindent
\caption{Calculated speed of sound using $c_s=\sqrt{\tilde{\gamma}/\rho\beta_T}$ as a function of MD steps for two NTP and NVE ensembles using the entire atoms in the MD simulation.
In normal water at the simulated pressure, the speed of sound is calculated close to $1600 m/s$ in the NPT ensemble.
With an increase of MD steps in normal water (before the formation of TS) the numerical value slowly converges to $1450 m/s$ consistent with the reported values.
We truncated the simulation at $1600 m/s$ to speed up the simulations.
The value of $c_s$ after TS formation is close to SW velocity calculated from the rate of nano-cavity radius.
}
\label{Fig10x}
\end{center}\vspace{-0.5cm}
\end{figure}

Finally, we combine all relevant thermodynamic variables to calculate the speed of sound as a local discontinuity due to the formation of TS.
As seen in Fig. \ref{Fig10x}, after initial fluctuations, the speed of water converges slowly to a correct value of $\approx 1450$ m/s.
Once the TS was added the speed of sound of mixed phases drops to that value and converges to  $\approx 1200$ m/s.

%%%%%%%%%%%%%%%%%%%%%%%%%%%%%%%%%%%%%%%%%%%%%%%%%%%%%%%
\subsection{Chemical changes}
Depending upon the initial temperature of TS, it may cause changes in the local chemical composition of the track and its surrounding environment.

Dissociation of $\ce{^{.}OH}-\ce{^{.}H}$ bond of a water molecule requires 118.8 kcal/mol equivalent of $\Delta T=$ 59,776 K. Whereas electron ionization of water molecule requires $\approx$ 13 eV,  equivalent of 150,858 K in the core of TS's.
% https://en.wikipedia.org/wiki/Bond-dissociation_energy

Thus isolated TS's with high enough kinetic energy, presumably generated by high LET charged particles (heavier elements than C, such as Fe) may act alone as a source of ionizing radiation.
Due to strong thermal fluctuations, bounded electrons and/or atoms to water molecules vibrate severely, overcome the potential energy barriers, and go over a transition state that turns the constituted electrons and/or atoms into unbounded from their host and form a local plasma.
A dissociation mechanism, similar to the photoelectric and/or Compton effect in the presence of $\gamma$-rays.

\begin{figure}
\begin{center}
\includegraphics[width=1.0\linewidth]{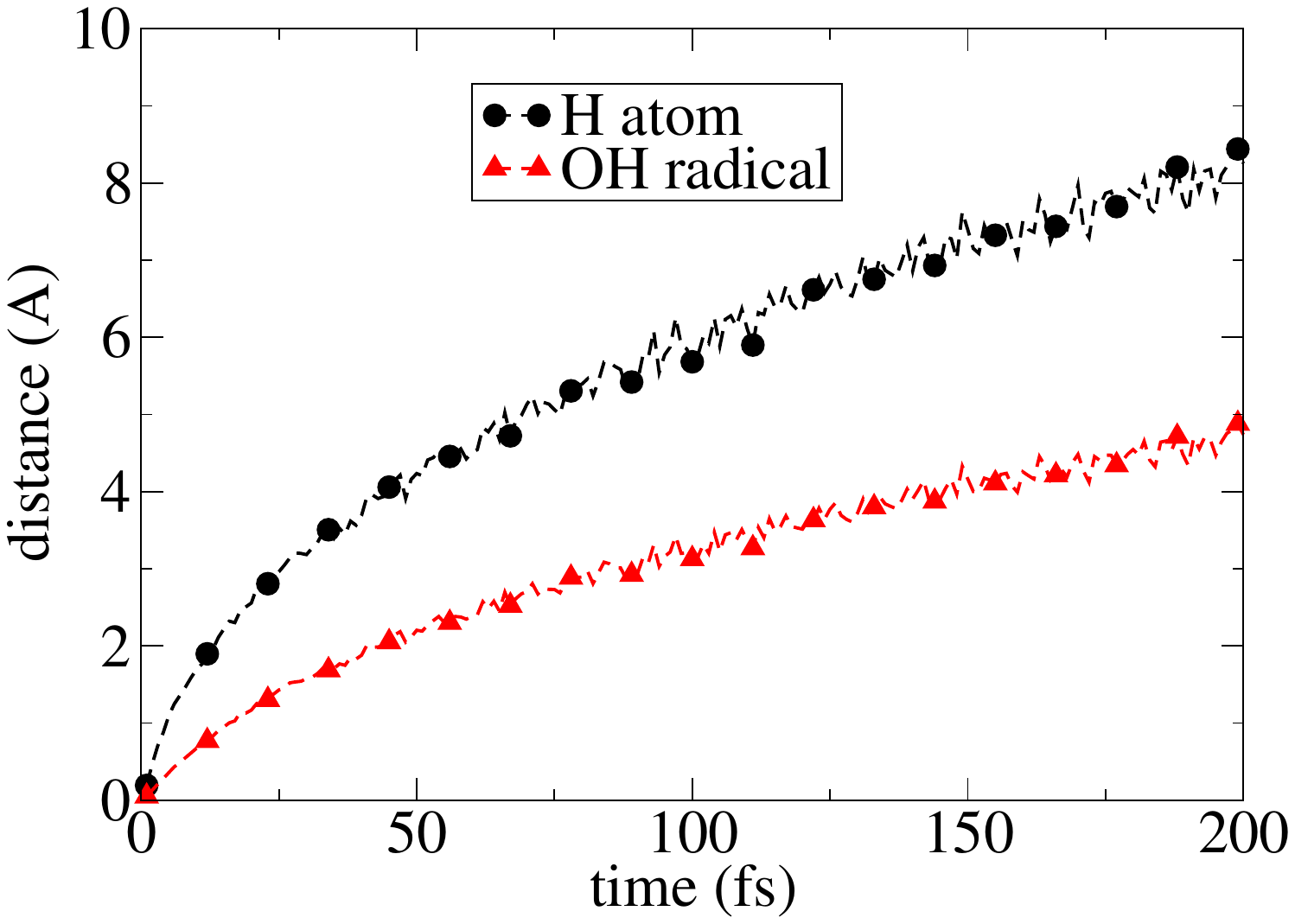}\\ %\vspace{-0.5cm} ---> Fig9
\noindent
\caption{Time-evolution of H-OH bond distance, corresponding to water molecules in SW ring/layer of nano-cavity.
The initial compression of covalent bonds due to the propagation of highly pressured shock-front stores a large potential energy.
Release of stored energy into kinetic energy of H dissociates water molecule and forms an atomistic H and \ce{^{.}OH}-radical.
This simulation demonstrates that the nano-cavity is not only mechanically active and its shock-front may destroy the biomolecules but it is also chemically active and it is full of ROS.
}
\label{Fig8}
\end{center}\vspace{-0.5cm}
\end{figure}

\begin{figure}
\begin{center}
\includegraphics[width=1.0\linewidth]{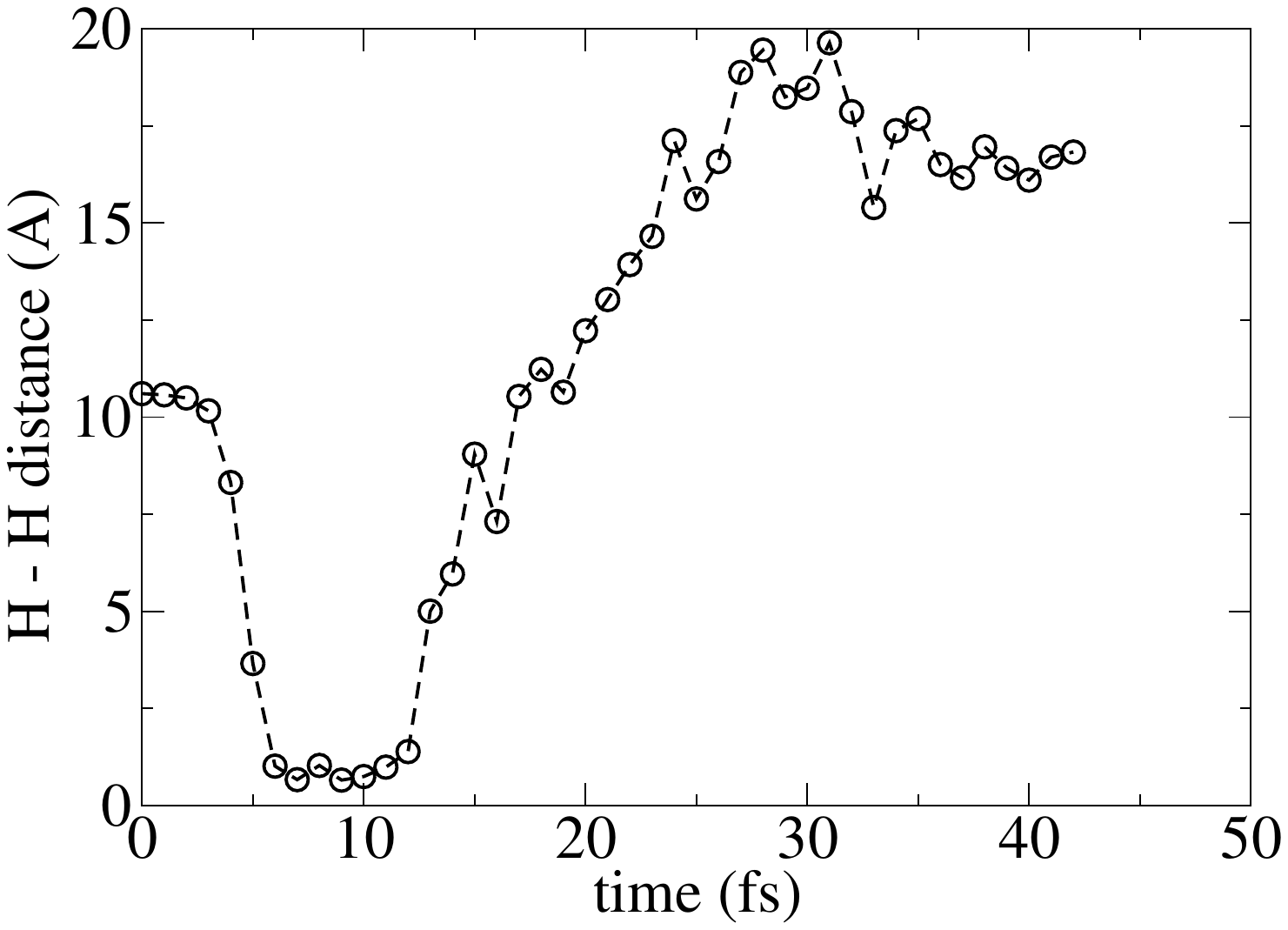}\\ %\vspace{-0.5cm} ---> Fig10
\noindent
\caption{Time-evolution of two randomly selected hydrogen atoms separated from their host water molecules in the nano-cavity ring. During MD simulation, they come close to each other, form a hydrogen atom, and become separated due to high-energy collisions by other molecules.
}
\label{Fig9}
\end{center}\vspace{-0.5cm}
\end{figure}

\begin{figure}
\begin{center}
\includegraphics[width=1.0\linewidth]{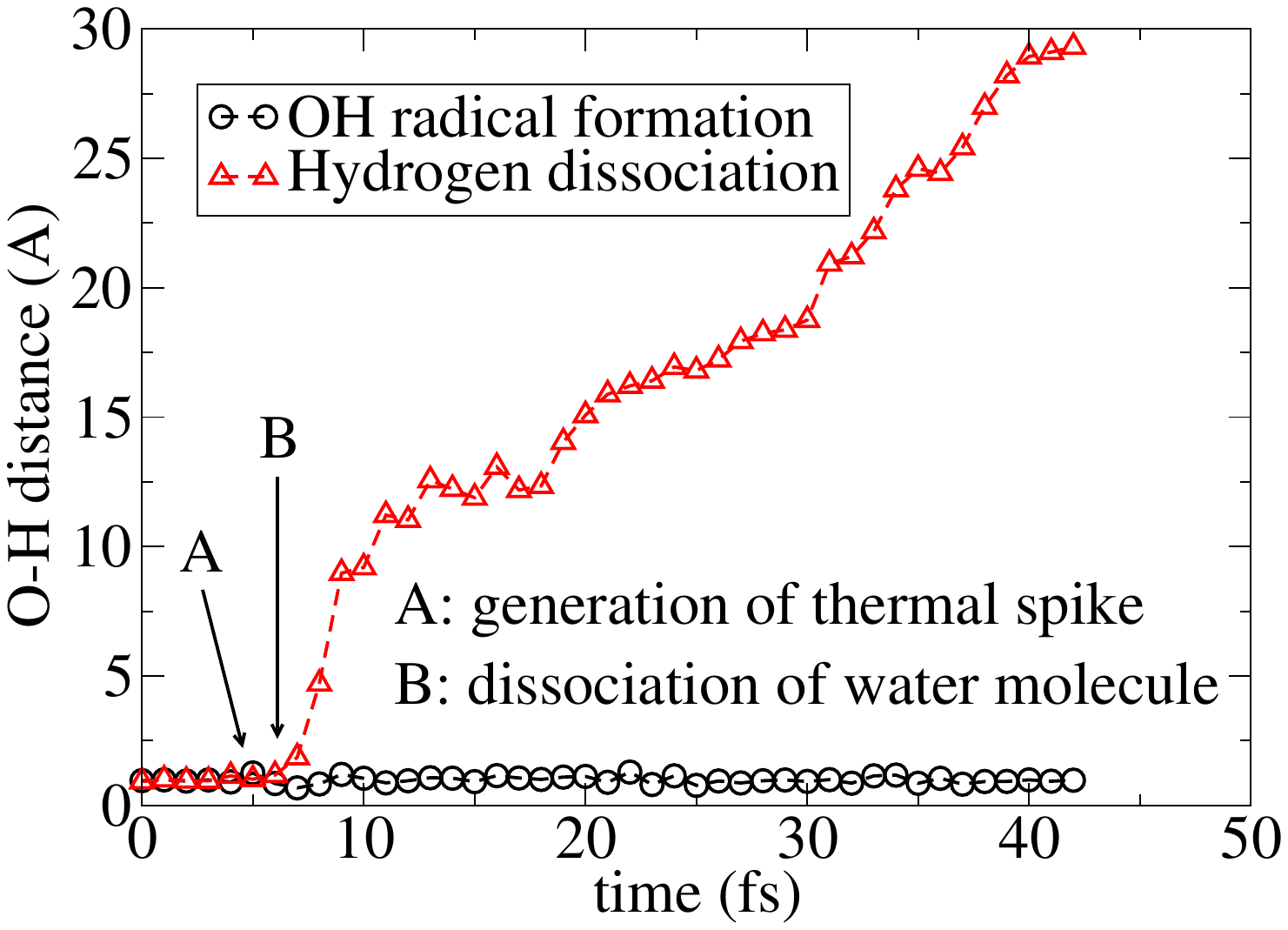}\\ %\vspace{-1.5cm} % ---> Fig11
\noindent
\caption{Time-evolution of a randomly selected water molecule in the ring of nano-cavity.
Due to highly energetic collisions hydrogen dissociates from the host and leaves an \ce{^{.}OH}-radical behind.
}
\label{Fig10}
\end{center}\vspace{-0.5cm}
\end{figure}

To illustrate this, during our MD simulations, we have picked isolated atoms and singled out events in which we have observed the separation of hydrogen atoms, hence the formation of \ce{^{.}OH}-radicals, incorporated with the transient formation and subsequently dissociation of $H_2$ molecules as a function of time, as shown in Figs. \ref{Fig8}-\ref{Fig10}.

%%%%%%%%%%%%%%%%%%%%%%%%%%%%%%%%%%%%%%%%%%%%%%%%%%%%%%%

\begin{figure}
\begin{center}
\includegraphics[width=1.0\linewidth]{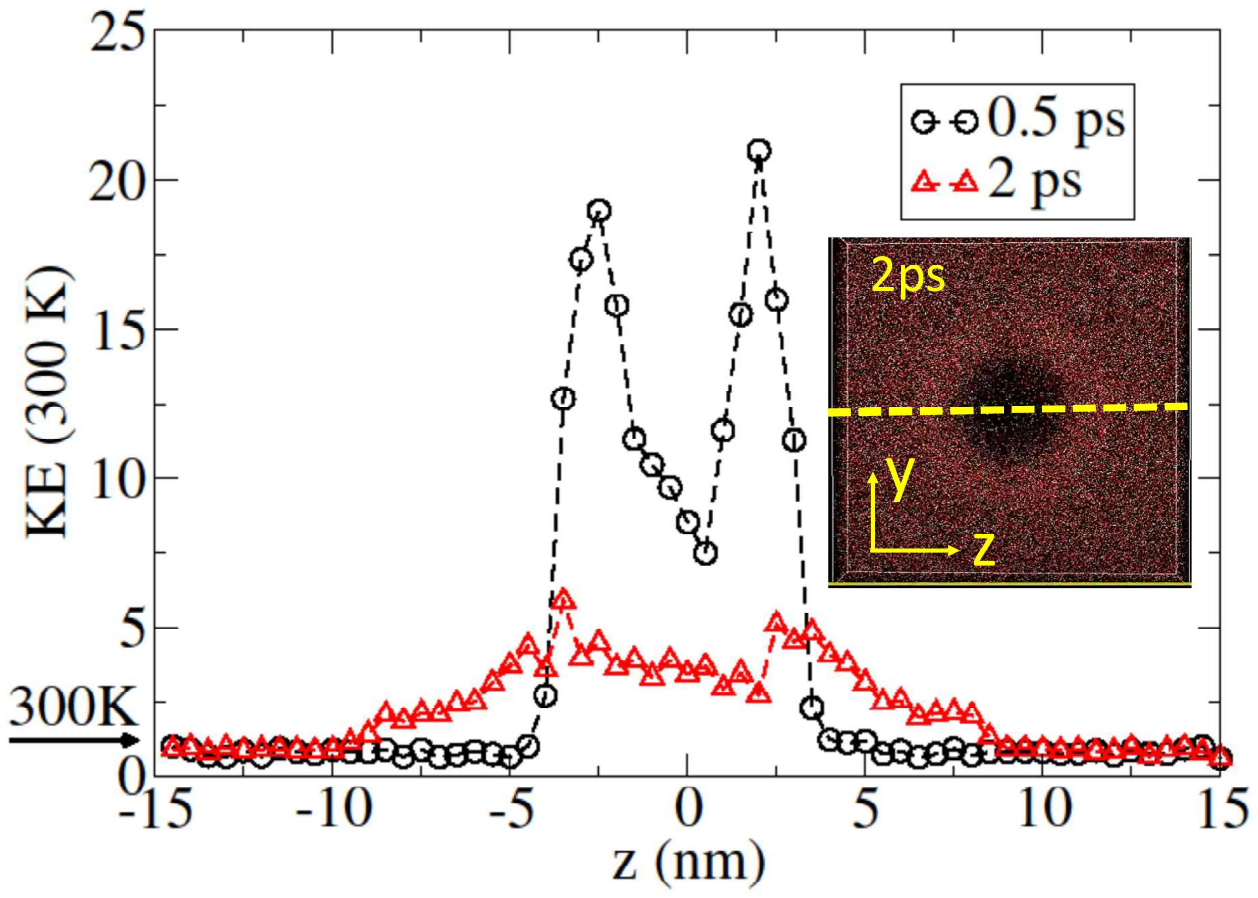}\\ \vspace{-2.5cm} % ---> Fig12
\noindent
\caption{Kinetic energy (KE) profile of water molecules scored in $0.5\times 0.5$ nm bins along the dashed line as depicted in the inset, corresponding to $t=0.5$ ps and $t=2$ ps of MD simulation in Fig. \ref{Fig3}.
The cusps in the KE are reminiscent of the compressed ring in the density of water as shown in the inset and in Fig. \ref{Fig3}.
The gaps in KE, behind and in front of the shoulders (inside and outside of the cylinder) can be interpreted as discontinuities in the local thermodynamic parameters.
The decay in the amplitude of the gap as a function of time which is an indication of SW crossing to a perturbative thermo-acoustic wave is visible.
The displacement of the cusps as a function of time provides a rough estimate of the SW front speed, $v_s = 5,000$ m/s.
}
\label{Fig4}
\end{center}\vspace{-0.5cm}
\end{figure}

\begin{figure}
\begin{center}
\includegraphics[width=1.0\linewidth]{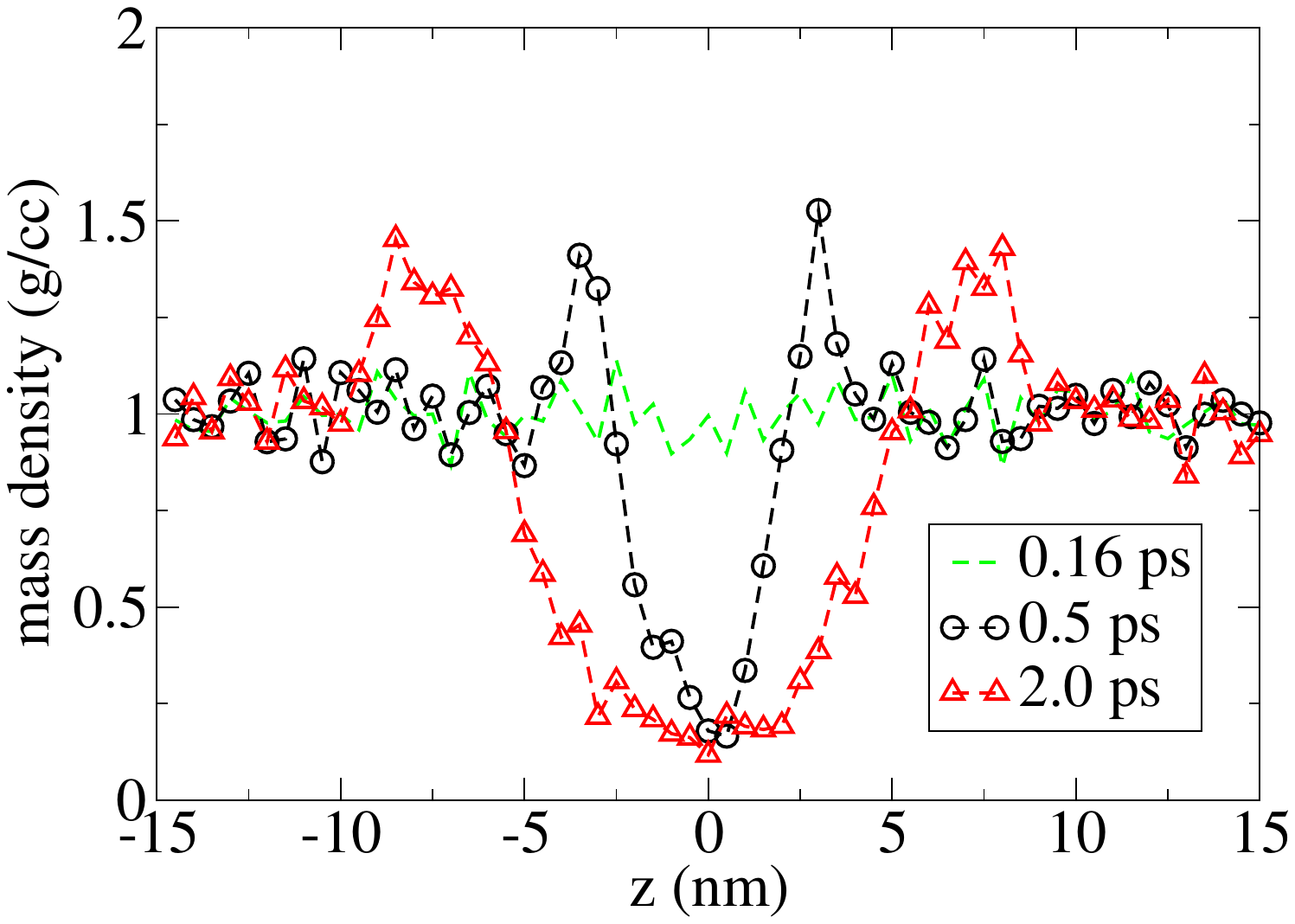}\\ %\vspace{-2.5cm} %---> Fig13
\noindent
\caption{Mass density profile of water molecules scored as in Fig. \ref{Fig4}.
The initial mean density of water molecules after thermal equilibration (up to 0.15 ps), fluctuates around the bulk mass density (green dashed line) even slightly after initiation of the TS.
It shows a time delay in response of molecular density to the rise in temperature and pressure of the TS.
A drop in density in the core of SW is a result of TS expansion in the form of SW.
}
% https://physics.stackexchange.com/questions/531829/what-s-the-difference-between-contact-discontinuity-and-shock-discontinuity
\label{Fig5}
\end{center}\vspace{-0.5cm}
\end{figure}

%%%%%%%%%%%%%%%%%%%%%%%%%%%%%%%%%%%%%%%%%%%%%%%%%%%%%%%%%%%%%%%%%%%
%%%%%%%%%%%%%%%%%%%%%%%%%%%%%%%%%%%%%%%%%%%%%%%%%%%%%%%%%%%%%%%%%%%
%%%%%%%%%%%%%%%%%%%%%%%%%%%%%%%%%%%%%%%%%%%%%%%%%%%%%%%%%%%%%%%%%%%
\begin{figure}
\begin{center}
\includegraphics[width=1.0\linewidth]{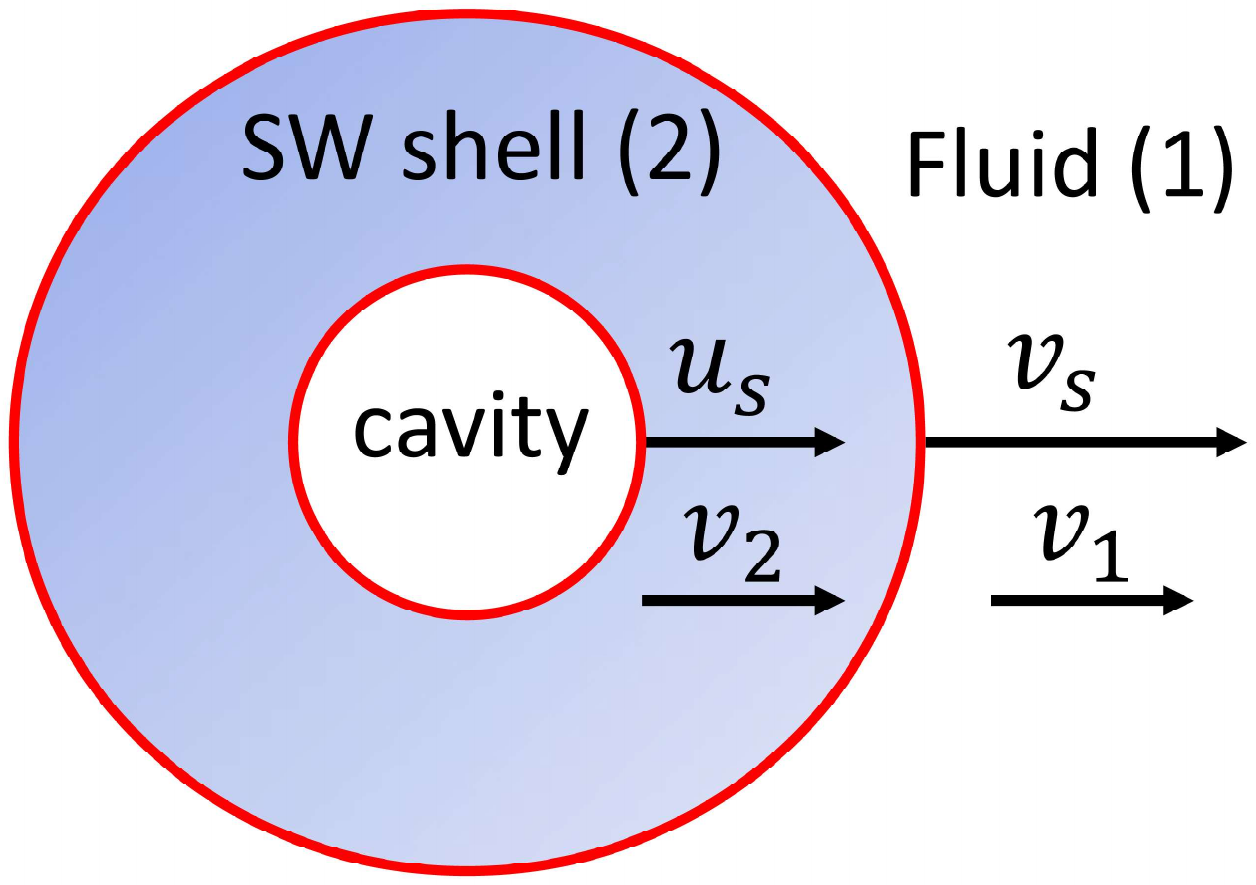}\\ \vspace{-2.5cm} % ---> Fig14
\noindent
\caption{
Sketch of an ideal cavity inside a fluid formed by pushing out the particles outward.
The hole represents the cavity and the shaded ring represents a nano-scale shock-wave shell.
The cavity grows with a rate proportional to the TS expansion velocity, $u_s$, that determines the shell thickness and the SW speed, $v_s$.
The velocities are given in the Lab frame of reference.
%The SW rest frame is the observer at rest on a co-moving tangent space of a cylinder.
}
\label{Fig2}
\end{center}\vspace{-0.5cm}
\end{figure}

\subsection{Nano cavities/bubbles}
In this section, we seek an extension of the analytical equations for the growth of the cavity separated by a wall thickness from the rest of the water molecules, consistent with the results obtained from MD simulations.

For the illustration of the problem, we have calculated a profile of KE and the mass density as a function of time as depicted in Figs.  \ref{Fig4} and \ref{Fig5}, using our MD simulation.
The cusps, interpolating the valleys inside of the cavity to the outside are the positions of the TS interface with a finite width.

An ideal cavity inside a fluid can be approximated with an empty space, like a void.
However, in the following section, we discuss a very rich internal structure of the cavity which exhibits oscillatory patterns in temperature, pressure, and density profiles.
As shown in Fig. \ref{Fig2}, at the interface between unperturbed fluid and cavity, internal pressure due to high temperature in the cavity pushes out the fluid particles outward.
Thus the cavity can be treated like a rigid cylinder with a radius growing outward with speed, $u_s$.
A rapid expansion of the cavity compresses the fluid in the shell, even if the unperturbed fluid is incompressible, like liquid water.

The compressed fluid forms a nano-meter size ring with a layer thickness which is determined by the in-flux of fluid particles and the speed of expansion, $u_s$.
The outer surface of the cylinder expands with a speed, higher than $u_s$, and can even reach a supersonic velocity at high temperatures, hence $v_s > c_s$.
%where $c_s$ is the speed of sound in the unperturbed fluid.
%Thus we use the acronym SW (shock-wave) to denote the outer layer of the cavity, whether or not it moves with a speed higher than $c_s$.

To analyze this ideal situation, we neglect the velocity of the fluid in region 1 (see Fig. \ref{Fig2}).
We further assume the compressed fluid within the TS shell moves with the cavity velocity, thus $v_1=0$ and $v_2 = u_s$.

In the rest frame of TS wavefront, where $v_s=0$, the radial velocities across the TS surface are transformed as $v_1 = -v_s$ and $v_2 = u_s - v_s = u_s + v_1$.
This corresponds to an observer sitting at the outer surface of TS, as sketched in Fig. \ref{Fig2}.
In terms of the conserved mass-flux density, $J$, the velocities can be rearranged as $v_1 = J V_1$, $v_2 = J V_2$ where $V_1 = 1/\rho_1$, $V_2 = 1/\rho_2$ are the corresponding specific volumes.
The conservation of momentum across the outer cavity interface in this TS frame of reference can be written
\begin{eqnarray}
P_1 + J^2 V_1 = P_2 + J^2 V_2,
\label{eq13}
\end{eqnarray}
which gives
\begin{eqnarray}
J = \pm \sqrt{\frac{P_2 - P_1}{V_1 - V_2}}.
\label{eq14c}
\end{eqnarray}
Considering the minus sign in Eq. (\ref{eq14c}) for $J$ which describes the flow of particles from 1 to 2 across the cavity interface as it grows, yields
\begin{eqnarray}
- u_s &=& v_1 - v_2 = J (V_1 - V_2) = - \sqrt{(P_2 - P_1)(V_1 - V_2)}
\nonumber \\ &&
= - \sqrt{\left(\frac{\rho_2 - \rho_1}{\rho_1\rho_2}\right) (P_2 - P_1)}.
\label{eq15}
\end{eqnarray}
Knowing that in the rest frame $v_1 > v_2$, we expect the layer thickness of TS shell to grow up
as a function of time.
This is due to the positive influx rate of the particles, accumulated behind the cavity interface in region 2, entering from the interface front in region 1.

The conservation of mass, momentum, and energy flux leads to
\begin{eqnarray}
\frac{\rho_1}{\rho_2} = \frac{v_2}{v_1} = \frac{u_s + v_1}{v_1} =
\frac{P_1(\tilde{\gamma}+1)+P_2(\tilde{\gamma}-1)}{P_1(\tilde{\gamma}-1)+P_2(\tilde{\gamma}+1)},
\label{eq16ss}
\end{eqnarray}
in which it can be algebraically manipulated to give
\begin{eqnarray}
\frac{P_2}{P_1} &=& 1 + \frac{\tilde{\gamma}(\tilde{\gamma}+1)}{4} \left(\frac{u_s}{\tilde{c}_s}\right)^2 \nonumber \\
&+& \tilde{\gamma}\left(\frac{u_s}{\tilde{c}_s}\right)\sqrt{1+\frac{(\tilde{\gamma}+1)^2}{16}\left(\frac{u_s}{\tilde{c}_s}\right)^2}.
\label{eqxxx}
\end{eqnarray}
With this ratio and with the help of Eq.(\ref{eq16ss}) it is straightforward to solve for $v_s$ in the Lab frame
\begin{eqnarray}
v_s = \frac{\tilde{\gamma}+1}{4} u_s + \tilde{c}_s\sqrt{1 + \frac{(\tilde{\gamma}+1)^2}{16}  \left(\frac{u_s}{\tilde{c}_s}\right)^2}.
\label{eqxxx_z}
\end{eqnarray}

Once the system reaches the equilibrium, the difference between the internal and external density and pressure vanishes, i.e., $\rho_2 = \rho_1$ and $P_2 = P_1$. Applying these conditions in Eqs. (\ref{eq16ss}-\ref{eqxxx_z}), we find $u_s = 0$, hence $v_s = \tilde{c}_s$. The SW front reduces to a regular sound wave that travels with speed $\tilde{c}_s$.

A similar equation to Eq. (\ref{eqxxx_z}) has been proposed empirically by Sing {\em et al.} [\onlinecite{Singh1980:PIAS}], such that
$v_s = a + b u_s$. Here $a = 1.364\times 10^{3}$ m/s and $b = 2.128$
where $a$ and $b$ are the fitting parameters obtained after interpolation to the experimental data.
At the limit of $u_s=0$, Sing {\em et al.} [\onlinecite{Singh1980:PIAS}] empirical results  predict $v_s = 1.364\times 10^{3}$ m/s, which is slightly below speed of sound in water.
Interestingly, in this limit our analytical Eq. (\ref{eqxxx_z}) predicts $v_s = a = \tilde{c}_s < c_s$ where $\tilde{c}_s = \sqrt{\tilde{\gamma} P/\rho}$, and $c_s = \sqrt{\tilde{\gamma}/\rho\beta_T}$. Recall that $c_s$ is the analytical expression for the speed of sound in water, calculated by our MD calculation (in the absence of TS), and was converged to 1450 m/s.
Moreover, as pointed out before, our MD simulation predicts $\tilde{\gamma} = 1.1$.
Comparing Sing {\em et al.}'s results with ours, $b = 2.128$ is almost four times higher than our analytical expression, $(\tilde{\gamma} + 1)/4 \approx 0.5$ for $u_s=0$, whereas in the limit of $u_s>>\tilde{c}_s$, where Eq.(\ref{eqxxx_z}) approaches to $v_s = u_s (\tilde{\gamma} + 1)/2 \approx u_s$, and our prediction gives, $b=1$, which is closer to the empirical value of the Sing {\em et al}. results.

\begin{figure}
\begin{center}
\includegraphics[width=1.0\linewidth]{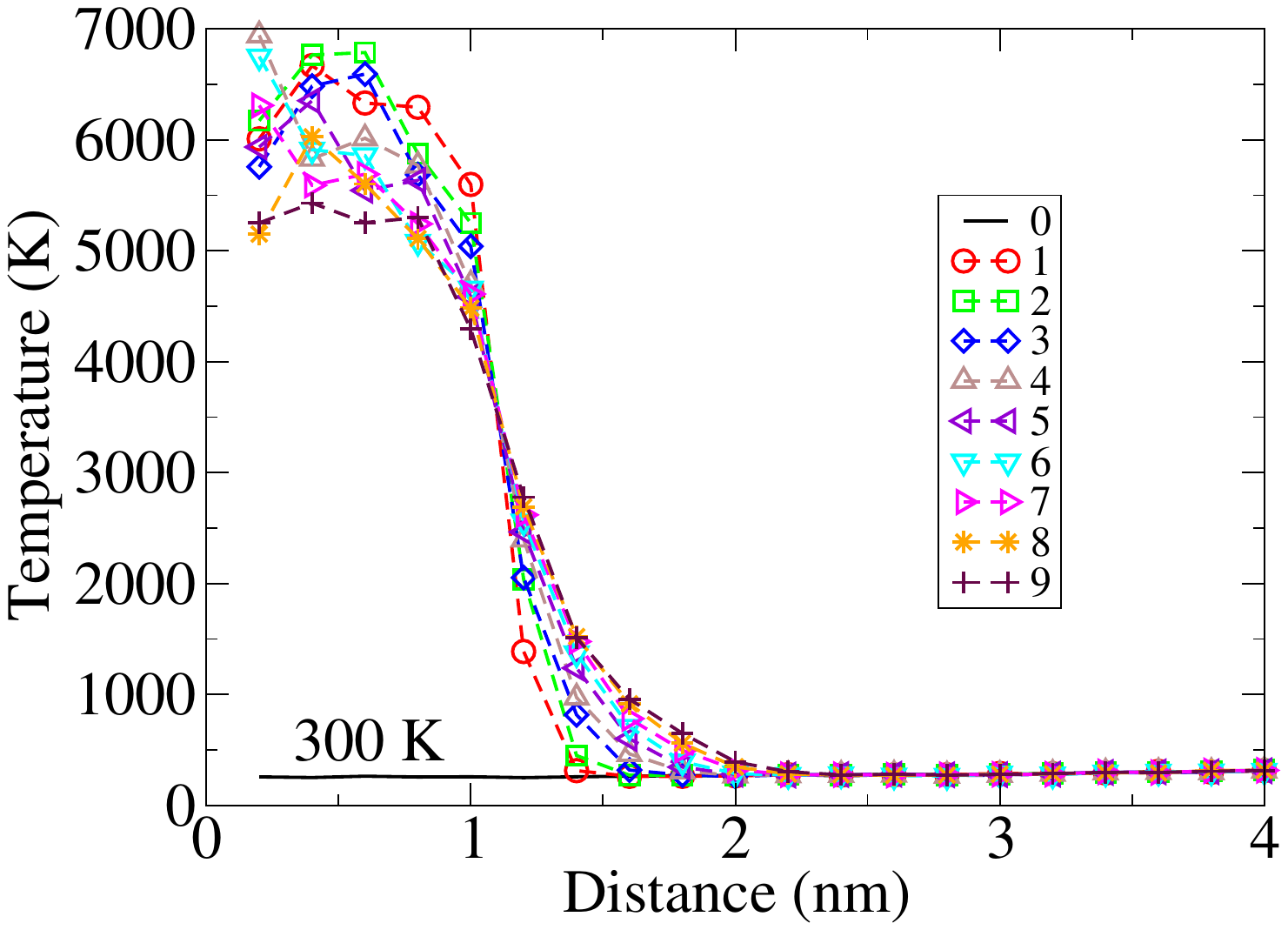}\\ \vspace{-0.4cm} % ---> Fig15
\includegraphics[width=1.0\linewidth]{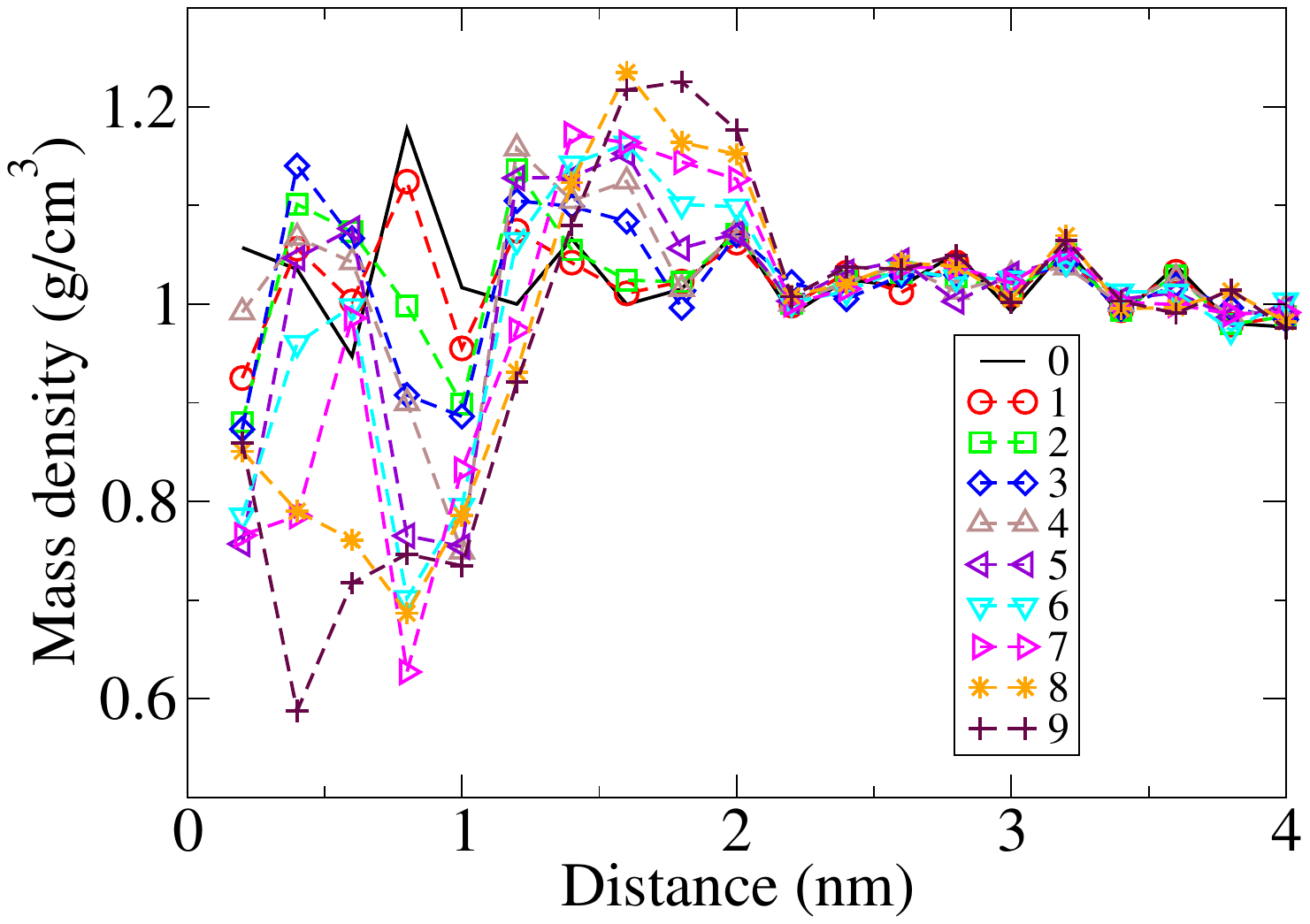}\\ \vspace{-0.4cm} % ---> Fig16
\includegraphics[width=1.0\linewidth]{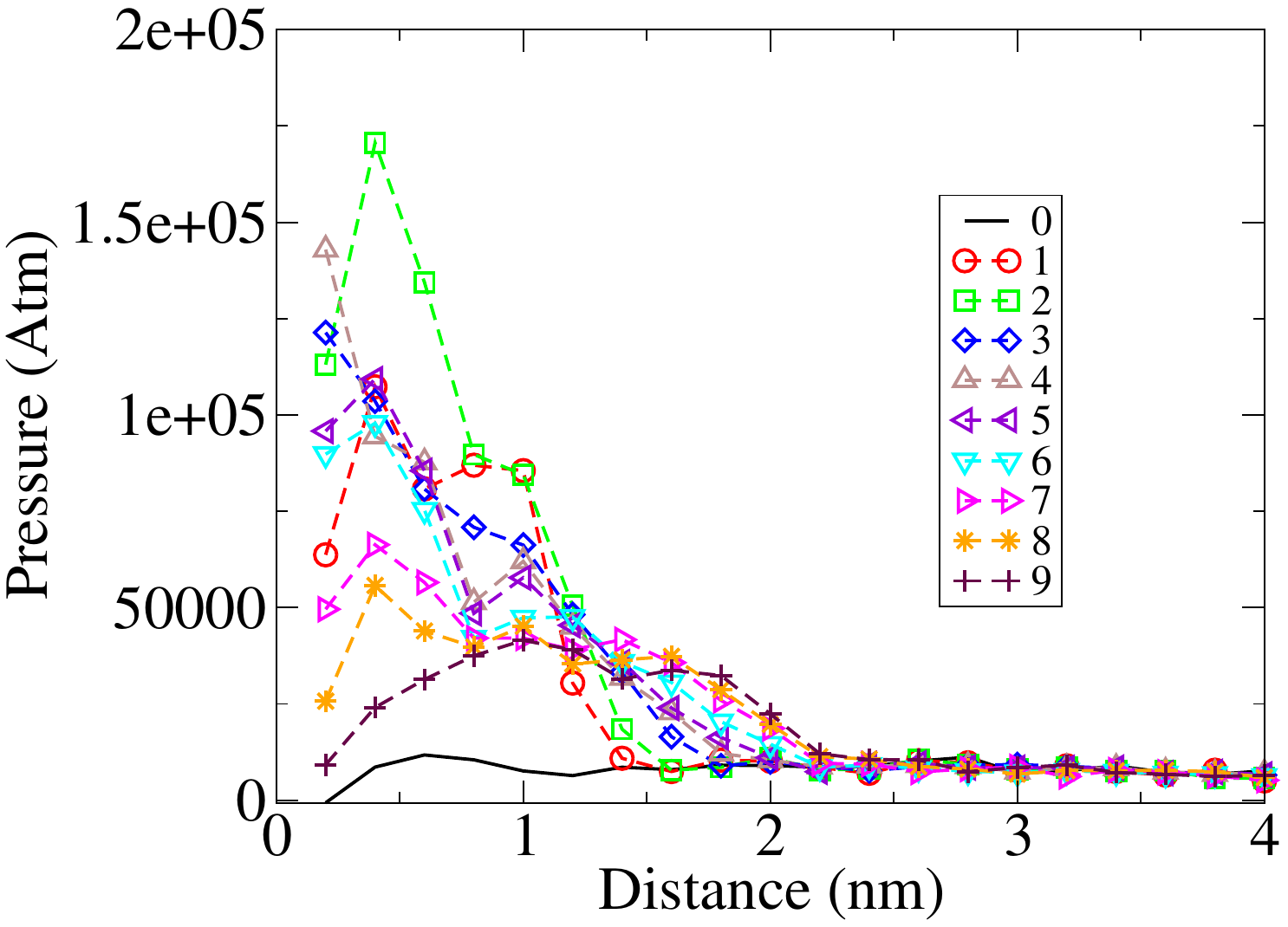}\\ \vspace{-0.5cm} % ---> Fig17
\noindent
\caption{Time-evolution of temperature, mass density, and pressure profiles in a series of concentric cylindrical bins. The initial TS was generated based on $^{12}$C parameters. Symbols corresponding to numbers are the sequence of the time. the black line corresponding to the symbol, 0, shows the state of water at the time before TS entrance. The oscillations in time in the first bin are visible. Across the interface, on the sixth bin, a sharp variation in temperature and mass density is seen in contrast to a smooth variation in pressure, an indication of a cross-over from shock to contact discontinuity.
The lateral thickness of bins is $\Delta r = 0.2$ nm. The longitudinal dimension of the bins spans over the entire 4 nm height of the computational box.
The time difference between sequential symbols corresponds is $\Delta t = 0.025$ ps thus
the last curve (shown by $+$), was sampled at $t=0.25$ ps after the generation of the TS.
The unit of pressure is the Atmosphere (1 Atm = 101.325 kPa).
}
\label{Fig11xxx}
\end{center}\vspace{-0.5cm}
\end{figure}

\subsection{Transition from shock to contact discontinuity}
As can be seen in Fig. \ref{Fig4}, the KE of the TS dissipates significantly within the first few ps after the creation of a TS.
However, the cusp in KE that separates inside and outside of the TS persists.
Similarly, the density difference between in and out of TS persists to exhibit a cusp, as seen in Fig. \ref{Fig5}.
However, one may expect the initial sharp difference in the pressure, inside and outside of the surface of discontinuity, to deform into a smooth curve.
We interpret this behavior as a persistent jump in the density and temperature at the surface of discontinuity but a smooth/continuous transition in the pressure difference,
a time evolution and a cross-over from shock to contact discontinuity
[\onlinecite{Landau_Lifshitz1987:Hydro_Book}].

Beyond this transition time, the thermal spike is expected to expand following standard heat transfer from a source of the circular symmetric high-temperature expanding cloud.
The mass transfer follows a similar diffusion equation [\onlinecite{Ljunggren1997:CS}] but with a time-varying diffusion constant that decays to the equilibrium value at room temperature.

%We postpone the details of our numerical analysis to our forthcoming publications.

To examine this hypothesis, we construct a slightly different type of binning using a series of concentric cylinders to sample pressure, temperature, and mass density as a function of time, as shown in Fig. \ref{Fig11xxx}.
Here 20 concentric cylindrical volumes from the center of the axis of the cylinder to a 4 nm radius, are used.
These volumes, which are stationary at the Lab frame of reference, span over the lateral dimension within the following range of $0 \leq r \leq 4$ nm. Thus $\Delta r = 0.2$ nm is the lateral thickness of the cylindrical shells.
The height of each bin spans over the entire height of the computational box, in this case, 4 nm.
As the thickness of each shell is fixed to 0.2 nm, the bin volume and the number of particles increase by the bin radius.

In Fig. \ref{Fig11xxx}, the symbols correspond to the time dependence scoring of the temperature, mass density, and pressure in each 100-time step, hence $\Delta t = 100\times 0.25$ fs $=0.025$ ps.
The black line with the symbol, 0, corresponds to the state of water, 0.025 ps before the initiation of the TS.
The last curve presented in Fig. \ref{Fig11xxx}, corresponding to the symbol ($+$), was sampled at $t=0.25$ ps after the generation of the TS.

The bin volume can be calculated by $L \times$ lateral area which is given by $\pi (r_2^2 - r_1^2) = \pi (r_2 - r_1) (r_2 + r_1) = \pi (4 {\rm nm}) (r_2 + r_1) \Delta r = 251.2 \AA^2 \times (r_2 + r_1)$. Here $r_1$ and $r_2$ are the inner and outer radius of the cylindrical bins.
For small $\Delta r = r_2 - r_1$, we can approximate $(r_2+r_1)$ to $2r$.
As the bin volume grows linearly with the cylinder radius, $r$, the number of particles grows linearly, given the constant density of water (1g/cm$^3$).
Hence, as it is visible in Fig. \ref{Fig11xxx}, the statistical fluctuations of the calculated variables lower down with an increase in $r$.

Evidently, within ps's time frame after the creation of the TS, during the formation of the nano-cavity, our analysis of temperature and pressure profiles reveals a novel internal periodic structure and an oscillatory pattern between low and high temperatures and pressures of the gas inside the cavity.
An internal periodic ``beating" structure of the gas motion, resembles the standing normal modes in a cylindrical acoustic pipe with expanding boundaries.
After induction of a large pressure inside the TS, the sudden expansion of the gas forms a local vacuum in the middle of the cavity.
Because the flow of the fluid goes along with a negative pressure gradient, the positive gradient in pressure, in the middle of the cavity pushes the escaped particles back to the center of the cavity. In this cycle,  a rise in temperature and pressure occurs once again.

Furthermore, because the direction of the gradient in pressure changes signs at the interface of the cavity, the stability of the cavity can be guaranteed due to the symmetry in pressure gradients with respect to the cavity interface, as seen in Fig. \ref{Fig11xxx}.
As a function of time, the curvature at the pressure peak, on the cavity interface, lowers and the slope approaches zero.
It is however possible that internal fluctuations disturb the balance between positive and negative pressure gradients.
If the internal positive pressure gradient prevails over the external negative pressure, the cavity collapses suddenly, and a phenomenon similar to the sonoluminescence effect takes place.
These fluctuations can be driven by classical and/or quantum zero-point oscillations of the electromagnetic field around the plasmon modes of the gas polarization in the cavity that breaks the symmetry of pressure gradients spontaneously, i.e., the Casimir effect [\onlinecite{Lamoreaux2005:RPP}].
Our hypothesis on the sudden collapse of TS nano-cavity may be considered as an alternative mechanism that explains the production of the Cherenkov light spectrum, with the emitted light intensity $\propto 1/\lambda$ (where $\lambda$ is the wavelength of the emitted light) during luminescence of water than prompt-$\gamma$ and emission of light
at lower energy than the Cherenkov light threshold hypothesized recently in Ref. [\onlinecite{Yamamoto2021:RPT}].

The transient state of water in the cavity at high temperatures and pressures but low densities is a manifestation of super-critical phase of water [\onlinecite{Zheng2020:EES}].
Thus we may also interpret the observed water luminescence as thermal radiation (e.g., black-body radiation, with gray-body corrections at longer wavelength in the optically thick regime) from the transient supercritical state of water localized in nanocavities wrapping around the track of charged particles.
We suggest the measurement of the spectrum, fitted to thermal radiation, where the intensity of emitted light scales like $1/\lambda^4$ for large $\lambda$'s may determine the average temperature of the spikes as a function of the charged particle penetration length, and  LET.

\subsection{Nano-bubble stability and inter-bubble interaction}
\begin{figure}
\begin{center}
\includegraphics[width=1.2\linewidth]{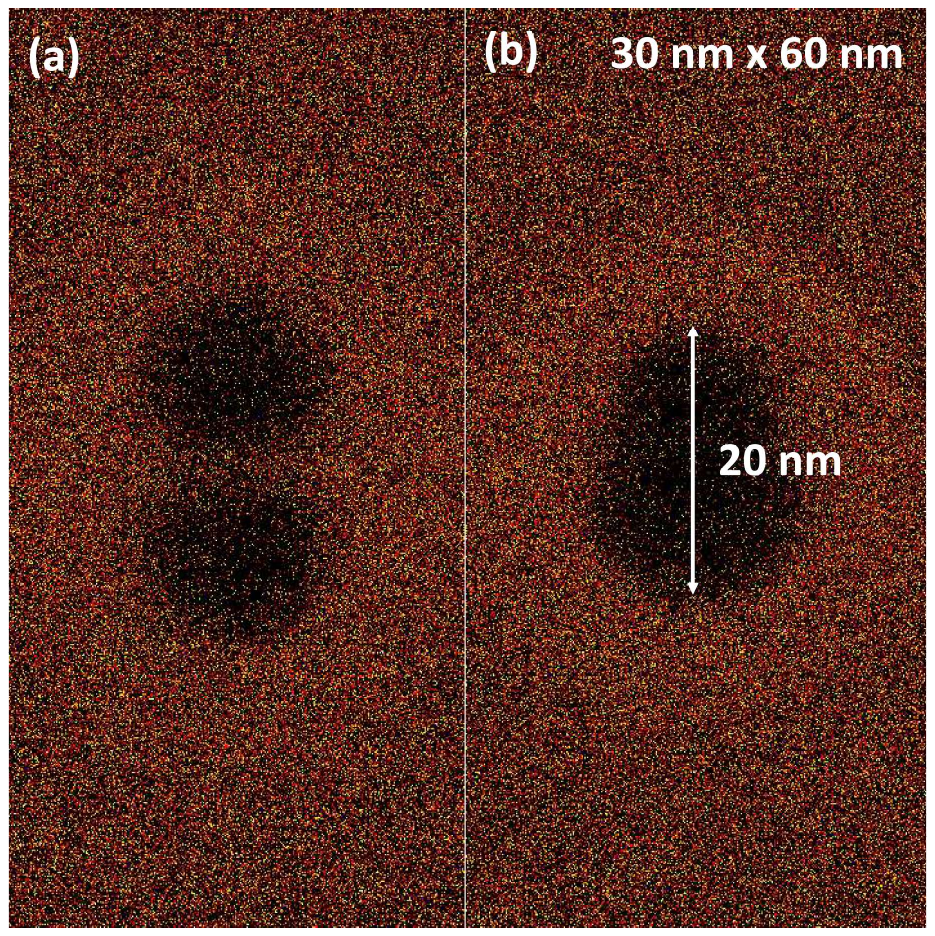}\\ \vspace{-2.5cm} % ---> Fig18
\noindent
\caption{Time-evolution of a system of two TS's created simultaneously (a) and with a time delay of 500 fs (b).
}
\label{Fig11}
\end{center}\vspace{-0.5cm}
\end{figure}

\begin{figure}
\begin{center}
\includegraphics[width=1.2\linewidth]{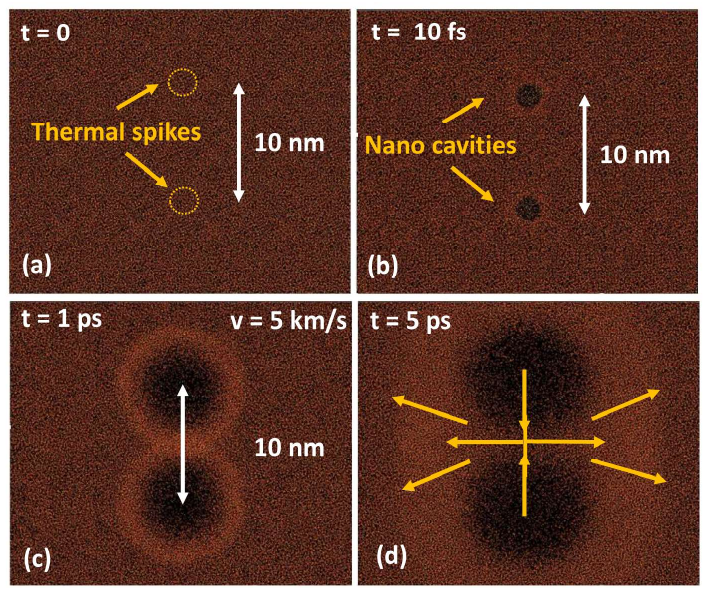}\\ \vspace{-2.5cm}  % ---> Fig19
\noindent
\caption{Time-evolution of a system of two TS's created simultaneously (a) and with a time delay of 500 fs (b).
}
\label{Fig12}
\end{center}\vspace{-0.5cm}
\end{figure}

% https://www.youtube.com/watch?v=lTGNbH2npME
To further examine the stability of the TSs discussed in the preceding section, we consider two TSs created in the vicinity of each other.
Upon their expansion, they may collide and form "molecular crowding" at their interface, one of the hypothetical track configurations at UHDRs as shown in Figs. \ref{Fig11} and \ref{Fig12}.

Fig. \ref{Fig11}(a) shows a snapshot of two thermal spikes created simultaneously 10 nm apart from each other at 2 ps after their creations, whereas in (b) there is a 0.5 ps time delay to their creation time.
A time delay beyond a threshold, which is a function of the TS's inter-separation distance and radius, allows penetration of the second TS into the first one.
In contrast, two TS's with time delay below the threshold, collide like two rigid objects with a high-density wall between them, like two stable bubbles touching each other at their interface.
The surface tension of the interface keeps the bubbles impenetrable.

\begin{figure}
\begin{center}
\includegraphics[width=1.2\linewidth]{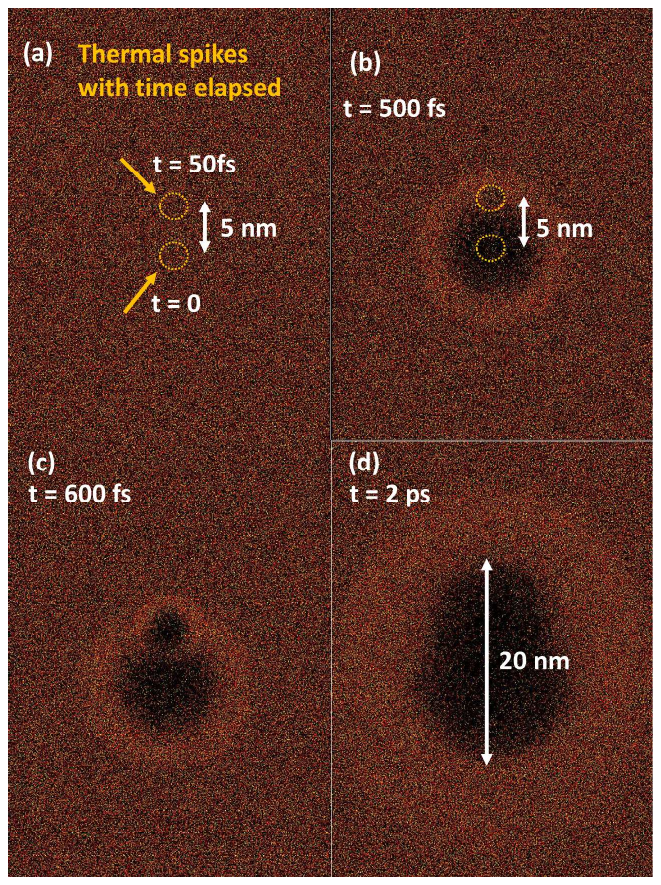}\\ \vspace{-2.5cm}  % ---> Fig20
\noindent
\caption{Time-evolution of a system of two TS's created with a time delay of 500 fs.
}
\label{Fig13}
\end{center}\vspace{-0.5cm}
\end{figure}

Fig. \ref{Fig13} shows the sequence of the selected events and how the time delay allows their penetration.
The circles in (a) are the initial location and dimension of the TS's.
The radius of each of them is 1 nm and the separation of 5 nm.
In (b) the first thermo-acoustic waveforms with a radius of 5 nm when the second TS enters the wall and expands (c) and finally they combine and form a large cavity with a diameter of 20 nm.

The flow of water molecules through the interface to outside TS's interfaces looks like a fountain-like and/or a jet stream with a side-wise direction to the outside of cavities.
The direction of the flows as shown by the arrows in Fig. \ref{Fig12} (d) forms a saddle point in the center. Such severe flow of water molecules makes them chemically active after they break apart through their covalent bonds losing electrons and leaving ions behind.

The creation of these two thermo-acoustic waves has been uploaded in the form of a video [\onlinecite{YouTube_two_SWs}] to help the readers visualize their time evolution and the inter-track couplings.
It shows the overall 5 ps running time.

%%%%%%%%%%%%%%%%%%%%%%%%%%%%%%%%%%%%%%%%%%%%%%%%%%%%%%%%%%%%%%%%%%%%%%%%%%
%%%%%%%%%%%%%%%%%%%%%%%%%%%%%%%%%%%%%%%%%%%%%%%%%%%%%%%%%%%%%%%%%%%%%%%%%%
%%%%%%%%%%%%%%%%%%%%%%%%%%%%%%%%%%%%%%%%%%%%%%%%%%%%%%%%%%%%%%%%%%%%%%%%%%
\subsection{Summary and remarks}
\label{SecRes}
TS's induced by charged particles generate a nano-scale cylindrical symmetric gradient in the temperature profile along the center of the track structure.
Assuming a one-to-one correspondence among the thermodynamic variables, specified by the equation of the state of the medium across the interface with finite width, $\ell$, a temperature gradient, $\vec{\nabla} T \approx (T_2 - T_1)/\ell$, produces a pressure gradient $\vec{\nabla} P  \approx (P_2 - P_1)/\ell$ normal to the cylindrical interface with volume, $\pi R^2 L$, with
the positive direction toward outside of this volume.
%This is analogous to a hedgehog field, normal to the surface of a single vortex with a topological winding number identical to the charge of a monopole in a 2D plane.

The location of the interface relative to the center of the TS, in the Lab frame of reference, is a dynamical variable and changes as a function of time with a characteristic rate given by $v_1$.
This is the speed of unperturbed fluid relative to TS surface of discontinuity, from the perspective of an observer stationary relative to the track interface.
In this system of coordinates where the Lab frame is moving in the opposite direction, $v_1$ represents the speed of the interface.
The location of the interface in the Lab frame, i.e., the radius of the cylinder can be calculated by $R(t) = \int_{0}^{t} dt' v_1(t')$.

Another important variable is the speed of sound in the unperturbed side of fluid which is given by the following equation $c_s = \sqrt{(C_P/C_V)/(\rho \beta_T)} = \sqrt{(C_P/C_V) (P/\rho) /(P\beta_T)}$.
%Here all thermodynamic variables are at point 1.
Note that $P\beta_T = 1$ for an ideal gas where $c_s = \sqrt{\tilde{\gamma} P/\rho}$ and $\tilde{\gamma} = C_P/C_V$ with numerical value corresponding to $c_s \approx 300 m/s$ in air at the standard condition.
For incompressible fluids $P\beta_T << 1$.
The inclusion of $P\beta_T$ in the equation for the speed of sound corrects and matches $c_s$ with their empirical values.
For water at room temperature, our MD calculation predicts the correct value of $c_s \approx 1450$ m/s.

For large enough gradients, $v_1$ may exceed the speed of sound.
We denote $M$, the Mach's number of the unperturbed fluids.
The supersonic flow of TS interfaces expands like shock waves (SW) of explosives with $M = v_1/c_s > 1$.

Our analysis has revealed that variabilities in the molecular mass and charge distribution and in general chemical, physical, and mechanical composition of the micro-environment of cells may dramatically alter the diffusion of \ce{^{.}OH}-radicals within nano-scales.

\section{Discussion and conclusion}
\label{SecConclu}
In this work, we have demonstrated the coexistence of a rapidly growing condensed state of water with a fast-growing hot spot, forming a stable state of water mixed with free radicals and chemical species generated due to bond breaking at high temperatures. The thin shell of a highly dense state of water grows by three to five folds of the speed of sound in water, forming a SW buffer, and wrapping around the nano-scale cylindrical symmetric cavity. The cavity consists of low density of water molecules, hence it forms a bubble-like state of water embedded in a room-temperature background of water. Eventually, SW amplitude dissipates to a sound wave. We have also studied the interaction of SWs and mapped the trajectory of water molecules into a jet stream along a vector field and identified a stable saddle point normal to the direction of SW propagation. We have demonstrated a random time-elapse mechanism among the TS's accounted for the nucleation, growth, and stability of the large bubbles.

The formation of supersonic growth of nano-bubbles / nano-cavities has been shown as an indication of a much higher diffusion constant currently predicted by the reaction-diffusion of the MC models in radiation physics. The current study is potentially significant in the accurate calculation of overlap integrals among the radiation tracks at ultra-high dose rates and the data analysis in FLASH radiotherapy that requires detailed knowledge of the spatial and temporal distribution of the particles in microscopic-scale pulsed beams. The more accurate simulation framework proposed hereby should help elucidate the intriguing mechanisms involved in FLASH radiotherapy. It may be used to define the dose fractionation requirements (i.e., instantaneous dose rate, irradiation time or beam scanning parameters) to maintain a FLASH effect while minimizing potential adverse effects [\onlinecite{Lascaud2023:PMB}]. Therefore, it may be a critical step toward safe clinical deployment of FLASH radiotherapy.

Recent efforts on developing of ionizing radiation acoustic imaging (iRAI) technique for real-time dosimetric measurements for FLASH radiotherapy [\onlinecite{Oraiqat2020:MP}] or more broadly radiotherapy with pulsed ionizing sources may also benefit from our simulation framework. Novel theoretical approaches may open up an in-depth understanding of the first principle mechanisms consistent with the picture explored in this work, i.e. starting from the formation of TSs and ending with the acoustic pulses used to reconstruct the dose in iRAI. It may foster the development of a new model-based reconstruction technique accounting for TS and nano-bubble generation to retrieve information not only on the dose distribution, as currently implemented, but also on the linear energy transfer (LET). Hence, it may unlock the \textit{in vivo} assessment of LET known to be more biologically relevant [\onlinecite{Abolfath2019:EPJD}] but which cannot be imaged with any methods currently available.

The new knowledge may also help improving various imaging techniques designed for the detection of charged particle tracks beyond radiotherapy, such as the optical emission of photons a mechanism known as scintillation in organic, inorganic, and in particular in liquid scintillation materials [\onlinecite{Beddar2016:Book}]. In addition to bubble formation in water in nuclear plants, NMR and MRI imaging beam topography [\onlinecite{Odeen2019:PNMRS,Gantza2023:PNAS}], and a many other applications in fundamental science and industry [\onlinecite{Winter2020:ANE,Brenner2002:RMP,Durante2011:RMP}].

Embedded in these results is a connection to the latest experiment on water luminescence, reported in [\onlinecite{Yamamoto2021:RPT}]. The authors have estimated the photon intensity by the Cherenkov light production from prompt gamma photons using MC simulations.
However, in this paper, we suggest an alternative mechanism. We propose that the indirect photon emission in the observed water luminescence may result either from the thermal radiation of the hot gas bubbles or the collapse of nano-bubbles because of strong fluctuations that drive the internal negative pressure gradient dominating the opposite pressure gradient outside of the bubble interface. One may speculate classical/quantum fluctuations that force bubbles (induced by high-energy charged particles) to collapse, to produce light, a mechanism similar to the Casimir effect [\onlinecite{Lamoreaux2005:RPP}] in the context of the sonoluminescence phenomenon [\onlinecite{Brenner2002:RMP}].
More characterization of the spectrum is needed to identify which channel of luminescence is responsible for the observation of the optical lights in the experiments [\onlinecite{Brenner2002:RMP}].
%A model in the literature related to classical and perhaps non-linear fluctuations that cause attractive force between two non-conducting objects, in this case, a thin layer of water surface that propagates with a supersonic speed. I am looking for such a theory.

In our model, for every charged particle passing through water, a nano-bubble is formed. They statistically collapse which results in photon emission from the location of the primary charged particles, hence their intensity is expected to be greater at the high LET regions, near the Bragg peak, and within the spots of low energy secondaries.
This mechanism can explain the tail of photon production from secondary protons beyond the range of the primary beam of the proton, distal to Bragg peak as illustrated in Ref. [\onlinecite{Yamamoto2021:RPT}].

Our simulations presented in this study have revealed a transient state of water in the cavity at high temperatures and pressures but low densities are formed within a radius of nano-meter from the charged particles. The calculated mass density profile of water inside these nano-cavities, shown in Figs. \ref{Fig5} and \ref{Fig11xxx}, matches with the thermodynamically stable state of supercritical water [\onlinecite{Zheng2020:EES}].
We therefore interpret the water luminescence observed and reported in Ref. [\onlinecite{Yamamoto2021:RPT}] as a manifestation of the thermal radiation from the transient supercritical state of water localized in nano-cavities wrapping around the track of charged particles.

\begin{acknowledgements}
The authors would like to thank useful scientific communications with Arash Darafsheh, Alberto Fraile, Jorge Kohanoff, Eugene Surdutovich, and Andrey Solov'yov.
The authors gratefully acknowledge the computational and data resources provided by the Leibniz Supercomputing Centre (www.lrz.de) and support from George Dedes.
NA is supported by the Natural Sciences and Engineering Research Council of Canada, the University of Waterloo, and the Perimeter Institute for Theoretical Physics.
Research at Perimeter Institute is supported in part by the Government of Canada through the Department of Innovation, Science and Economic Development Canada and by the Province of Ontario through the Ministry of Colleges and Universities. MR, KP, and JL acknowledge the support from the Centre for Advanced Laser Applications and LMUexcellent, funded by the Federal Ministry of Education and Research (BMBF) and the Free State of Bavaria under the Excellence Strategy of the Federal Government and the L\"ander.
\end{acknowledgements}

\noindent{\bf Authors contributions:}
RA: wrote the main manuscript, prepared figures, and performed mathematical derivations and computational steps. NA, SR, MR, RT, JL: contributed to developing scientific ideas, computational analysis of the problem, and writing the manuscript.
AvD: provided ReaxFF, data analysis, and wrote the manuscript.
KP, JL, and RM: wrote the main manuscript, proposed the scientific problem, , co-supervised the project, and provided critical feedback. 

\noindent{\bf Competing financial interest:}
The authors declare no competing financial interests.

\noindent{\bf Corresponding Author:      $~~~~~~~~~~~~~~~~~~~~~~~~~$}\\
$^\dagger$: ramin1.abolfath@gmail.com
%$^*$ rmohan@mdanderson.org

%%%%%%%%%%%%%%%%%%%%%%%%%%%%%%%%%%%%%%%%%%%%%%%%%%%%%%%%%%%%%%%%%%%%%%%%%%
%%%%%%%%%%%%%%%%%%%%%%%%%%%%%%%%%%%%%%%%%%%%%%%%%%%%%%%%%%%%%%%%%%%%%%%%%%


\begin{thebibliography}{99}
\bibitem{Norman1963:NSE}
A. Norman, P. Spiegler,
{\it Radiation Nucleation of Bubbles in Water}, Nucl. Sci. Eng. {\bf 16}, 213-217 (1963).

\bibitem{Vedadi2010:PRL}
M. Vedadi, A. Choubey, K. Nomura, R. K. Kalia, A. Nakano, P. Vashishta, and A. C. T. van Duin,
{\it Structure and Dynamics of Shock-Induced Nanobubble Collapse in Water}, Phys. Rev. Lett. {\bf 105}, 014503 (2010).

\bibitem{Min2019:JCP}
S. H. Min, M. L. Berkowitz,
{\it Bubbles in water under stretch-induced cavitation}, J. Chem. Phys. {\bf 150}, 054501 (2019).
%; doi: 10.1063/1.5079735

\bibitem{Mallory2016:CROH}
M. Mallory, E. Gogineni, G. C. Jones, L. Greer, C. B. Simone II,
{\it Therapeutic hyperthermia: The old, the new, and the upcoming},
Critic. Rev. Onc. Hemato.,  {\bf 97}, 56-64 (2016).
% https://doi.org/10.1016/j.critrevonc.2015.08.003.


\bibitem{Wan2016:CBM}
G.-Y. Wan, Y. Liu, B.-W. Chen, Y.-Y. Liu, Y.-S. Wang, N. Zhang,
{\it Recent advances of sonodynamic therapy in cancer treatment}
Cancer Biol Med 2016. doi: 10.20892/j.issn.2095-3941.2016.0068

\bibitem{Yan2020:CTRO}
W. Yan, M. K. Khan, X. Wu, C. B. Simone, J. Fan, E. Gressen, X. Zhang, C. L. Limoli, H. Bahig, S. Tubin, and W. F. Mourad, {\it Spatially fractionated radiation therapy: History, present and the future},
Clin. Transl. Radiat. Oncol. {\bf 20}, 30–38 (2020).


\bibitem{Favaudon2014:STM}
Favaudon V, Caplier L, Monceau V, et al. Ultrahigh dose-rate FLASH irradiation increases the differential response between normal and tumor tissue in mice. Sci Transl Med. 2014;6:1-9.
%Favaudon V, Caplier L, Monceau V, Pouzoulet F, Sayarath M, Fouillade C, Poupon MF, Brito I, Hupe P, Bourhis J, Hall J, Fontaine JJ, Vozenin MC, {\it Ultrahigh dose-rate FLASH irradiation increases the differential response between normal and tumor tissue in mice}, Sci Transl Med. {\bf 6}, 1-9 (2014). %doi: 10.1126/scitranslmed.3008973.

\bibitem{Montay-Gruel2018:RO}
Montay-Gruel P, Bouchet A, Jaccard M, et al. X-rays can trigger the FLASH effect: Ultra-high dose-rate synchrotron light source prevents normal brain injury after whole brain irradiation in mice. Radiother Oncol. 2018;129(3):582-588.
%Montay-Gruel P, Bouchet A, Jaccard M, Patin D, Serduc R, Aim W, et al. {\it X-rays can trigger the FLASH effect: Ultra-high dose-rate synchrotron light source prevents normal brain injury after whole brain irradiation in mice}, Radiother Oncol {\bf 129} 582–8 (2018).

\bibitem{Vozenin2018:CCR}
Vozenin MC, De Fornel P, Petersson K, et al. The advantage of FLASH radiotherapy confirmed in mini-pig and catcancer patients. Clin Cancer Res 2019 Jan 1;25(1):35-42.
%Vozenin MC, De Fornel P, Petersson K, Favaudon V, Jaccard M, Germond JF, et al. The advantage of Flash radiotherapy confirmed in mini-pig and catcancer patients. Clin Cancer Res 2018. https://doi.org/10.1158/1078-0432.CCR-17-3375.

\bibitem{Montay-Gruel2019:PNAS}
Montay-Gruel P, Acharya MM, Petersson K, et al. Long-term neurocognitive benefits of FLASH radiotherapy driven by reduced reactive oxygen species. Proc Natl Acad Sci USA.  2019;116(22):10943-10951.
%Montay-Gruel P, Acharya MM, Petersson K, Alikhani L, Yakkala C, Allen BD, Ollivier J, Petit B, Jorge PG, Syage AR, Nguyen TA, Baddour AAD, Lu C, Singh P, Moeckli R, Bochud F, Germond JF, Froidevaux P, Bailat C, Bourhis J, Vozenin MC, Limoli CL, {\it Long-term neurocognitive benefits of FLASH radiotherapy driven by reduced reactive oxygen species}, PNAS {\bf 116}, 10943–10951 (2019).

\bibitem{Buonanno2019:RO}
Buonanno M, Grilj V, Brenner DJ. Biological effects in normal cells exposed to FLASH dose rate protons. Radiother Oncol. 2019;139:51-55.
%Buonanno M, Grilj V, Brenner DJ, {\it Biological effects in normal cells exposed to FLASH dose rate protons}, Radiother Oncol {\bf 139} 51–55 (2019).

\bibitem{Vozenin2019:RO}
Vozenin MC, Baumann M, Coppes RP, Bourhis J. FLASH radiotherapy international workshop. Radiother Oncol. 2019;139:1-3.
%Vozenin MC, Baumann M, Coppes RP, Bourhis J, {\it FLASH radiotherapy International Workshop}, Radiother Oncol {\bf 139}, 1–3 (2019).

\bibitem{Arash2020:MP}
Darafsheh A, Hao Y, Zwart T, Wagner M, Catanzano D, Williamson JF, Knutson N, Sun B, Mutic S, Zhao T.
Feasibility of proton FLASH irradiation using a synchrocyclotron for preclinical studies.
Med Phys. 2020; doi: 10.1002/mp.14253.

\bibitem{Spitz2019:RO}
Spitz DR, Buettner GR, Petronek MS, et al. An integrated physico-chemical approach for explaining the differential impact of FLASH versus conventional dose rate irradiation on cancer and normal tissue responses. Radiother Oncol. 2019; 139:23-27.
%Spitz DR, Buettner GR, Petronek MS, St-Aubin JJ, Flynn RT, Waldron TJ, et al. {\it An integrated physico-chemical approach for explaining the differential impact of FLASH versus conventional dose rate irradiation on cancer and normal tissue}, responses. Radiother Oncol {\bf 139}, 23–7 (2019).

\bibitem{Koch2019:RO}
Koch CJ. Re: Differential impact of FLASH versus conventional dose rate irradiation. Radiother Oncol. 2019;139:62-63.
%Koch CJ, {\it Differential impact of FLASH versus conventional dose rate irradiation}, Radiother Oncol {\bf 139} 62–63 (2019).

\bibitem{Abolfath2020:MP}
R. Abolfath, D. Grosshans, R. Mohan, {\it Oxygen depletion in FLASH ultra-high-dose-rate radiotherapy: A molecular dynamics simulation}, Med. Phys. {\bf 47}, 6551-6561 (2020).

\bibitem{Seco2021:MP}
J. Jansen, J. Knoll, E. Beyreuther, J. Pawelke, R. Skuza, R. Hanley, S. Brons, F. Pagliari, J. Seco, {\it Does FLASH deplete oxygen? Experimental evaluation for photons, protons, and carbon ions}, Med. Phys. {\bf 48}, 3982 (2021).

\bibitem{Abolfath2023:FP}
R. Abolfath, A. Baikalov, S. Bartzsch, E. Schuler, R. Mohan, {\it A
stochastic reaction-diffusion modeling investigation of FLASH ultra-high dose rate
response in different tissues, Special issue on Multidisciplinary Approaches to the
FLASH radiotherapy}, Front. Phys., {\bf 11}, 1060910 (2023).
%DOI 10.3389/fphy.2023.1060910

\bibitem{Baikalov2023:FP}
A. Baikalov, R. Abolfath, E. Schuler, R. Mohan, Jan Wilkens, S. Bartzsch, {\it A
Intertrack interaction at ultra-high dose rates and its role in the FLASH effect},
Front. Phys. {\bf 11}, 1215422 (2023).
%doi: 10.3389/fphy.2023.1215422

\bibitem{Oraiqat2020:MP}
I. Oraiqat, W. Zhang, D. Litzenberg, K. Lam, N. B. Sunbul, J. Moran, K. Cuneo, P. Carson, X. Wang, I. E. Naqa,
{\it An ionizing radiation acoustic imaging (iRAI) technique for real-time dosimetric measurements for FLASH radiotherapy},
Med. Phys. {\bf 47}, 5090-5101 (2020).

\bibitem{Kalinichenko2001:chapterbook}
A.I. Kalinichenko, V.T. Lazurik, and I.I. Zalyubovsky, The Physics and Technology of Particle and Photon Beams Volume 9: Introduction to Radiation Acoustics. The Netherlands: Harwood Academic Publishers, 2001.

\bibitem{Hickling2018:MP}
S. Hickling, L. Xiang, K. C. Jones, K. Parodi, W. Assmann, S. Avery, M. Hobson, I. El Naqa, {\it Ionizing radiation-induced acoustics}, Med. Phys. {\bf 45}, e707 (2018).

\bibitem{Lascaud2023:PMB}
J. Lascaud and K. Parodi,
{\it On the potential biological impact of radiation-induced acoustic emissions during ultra-high dose rate electron radiotherapy: a preliminary study}, Phys. Med. Biol. {\bf 68}, 05LT01 (2023).

\bibitem{Beddar2016:Book}
A. S. Beddar, and L Beaulieu, {\it Scintillation Dosimetry}. Boca Raton, FL: CRC
Press (2016).

\bibitem{Odeen2019:PNMRS}
H. Odeen, D. L. Parker, {\it Magnetic resonance thermometry and its biological applications – Physical principles and practical considerations}, Prog. Nucl. Magn. Reson. Spectrosc. {\bf 110} 34–61 (2019).
%doi:10.1016/j.pnmrs.2019.01.003.

\bibitem{Gantza2023:PNAS}
S. Gantza, L. Karscha, J. Pawelkea , J. Petera, S. Schellhammer, J. Smeets, Erik van der Kraaij, A. Hoffmann,
{\it Direct visualization of proton beam irradiation effects in liquids
by MRI}, PNAS {\bf 120}, 23 e2301160120 (2023).
% https://doi.org/10.1073/pnas.2301160120

\bibitem{Winter2020:ANE}
G. E. Winter, C. M. Cooling, M. D. Eaton,
{\it Linear energy transfer of fission fragments of 235U and nucleation of gas bubbles in aqueous solutions of uranyl nitrate}
Annals of Nuclear Energy {\bf 142}, 107379 (2020).
%https://doi.org/10.1016/j.anucene.2020.107379.

\bibitem{Brenner2002:RMP}
M. P. Brenner, S. Hilgenfeldt, D. Lohse,
{\it Single-bubble sonoluminescence}, Rev. Mod. Phys., {\bf 74}, 2, (2002).

\bibitem{Durante2011:RMP}
M. Durante, F. A. Cucinotta,
{\it Physical basis of radiation protection in space travel}, Rev. Mod. Phys. {\bf 83}, 1245 (2011).

\bibitem{Abolfath2022:PMB}
R. Abolfath, A. Baikalov, S. Bartzsch, N. Afshordi, and R. Mohan, “The effect of non-ionizing excitations
on the diffusion of ion species and inter-track correlations in flash ultra-high dose rate radiotherapy,”
Phys. Med. Biol. {\bf 67}, 105 (2022).

\bibitem{Kellerer1985:chapterbook}
A. M. Kellerer, {\it Fundamentals of Microdosimetry the Dosimetry of Ionizing
Radiation}. Vol. 1, Kase KR, et al. London: Academic, 77–161 (1985).

\bibitem{Fain1974:RR}
J. Fain, M. Monnin, M. Montret, {\it Spatial Energy Distribution Around Heavy-Ion Path}, Radiat. Res. 57, 379-389 (1974).

\bibitem{Wang2014:PMB}
H. Wang, O. N Vassiliev, {\it Radial dose distributions from protons of therapeutic energies calculated with Geant4-DNA}, Phys. Med. Biol. 59 (2014) 3657–3668.

\bibitem{LaVerne2000:RR}
J. A. LaVerne,
{\it Track Effects of Heavy Ions in Liquid Water}, Radiat. Res. {\bf 153}, 487–496 (2000).

\bibitem{Agostinelli2003:NIMA}
S. Agostinelli {\em et. al.}, Nucl. Instrum. Meth. A {\bf 506}, 250 (2003).

\bibitem{Incerti2010:IJMSSC}
Incerti S. {\em et. al.}, {\it The GEANT4-DNA project}, Int. J. Modelling Simul. Sci. Comput. {\bf 1}, 157–78 (2010).

\bibitem{vanDuin2001:JPCA}
van Duin, A. C. T.; Dasgupta, S.; Lorant, F.; Goddard, W. A.,
{\it ReaxFF: A Reactive Force Field for Hydrocarbons}, J. Phys. Chem. A 2001, 105, 9396.

\bibitem{Yusupov2012:NJP}
M Yusupov, E C Neyts, U Khalilov, R Snoeckx, A C T van Duin, and A Bogaerts,
{\it Atomic-scale simulations of reactive oxygen plasma species interacting with bacterial cell walls},
New Journal of Physics 14 (2012) 093043.

\bibitem{Verlackt2015:NJP}
Verlackt, C.C., Neyts, E.C., Jacob, T., Fantauzzi, D., Golkaram, M., Shin, Y.K., van Duin, A.C.T. and Bogaerts, A. (2015) Atomic scale pathways in DNA oxidation by hydroxyl radicals: a reactive molecular dynamics study for plasma oncology. New Journal of Physics 17, 103005-103001.

\bibitem{Abolfath2011:JPC}
R. M. Abolfath, A. C. T. van Duin, T. Brabec, {it Reactive molecular dynamics study on the first steps of DNA damage by free hydroxyl radicals}, J. Phys. Chem. A {\bf 115}, 11045 (2011).
See the real-time simulations and movies at:
http://qmsimulator.wordpress.com/

\bibitem{deVera2019:CN}
P. de Vera, E. Surdutovich, A. V. Solov’yov, {\it The role of shock waves on the biodamage induced by ion beam radiation}, Cancer Nano {\bf 10}, 5, (2019). https://doi.org/10.1186/s12645-019-0050-3

\bibitem{deVera2018:EPJD}
P. de Vera, E. Surdutovich, N. J. Mason, F. J. Currell, A. V. Solov'yov,
{\it Simulation of the ion-induced shock waves effects on the transport of chemically reactive species in ion tracks}
Eur. Phys. J. D {\bf 72}, 147 (2018).

\bibitem{Surdutovich2010:PRE}
E. Surdutovich, A. V. Solov’yov, {\it Shock wave initiated by an ion passing through liquid water}, Phys Rev E. {\bf 82}, 051915 (2010).

\bibitem{Surdutovich2014:EPJD}
E. Surdutovich, A.V. Solov’yov,
{\it Multiscale approach to the physics of radiation damage with ions}, Eur. Phys. J. D {\bf 68}, 353 (2014).

\bibitem{Fraile2019:JCP}
A. Fraile, M. Smyth,  J. Kohanoff, A. V. Solov’yov, {\it First principles simulation of damage to solvated nucleotides due to shock waves}, J. Chem. Phys. {\bf 150}, 015101 (2019); https://doi.org/10.1063/1.5028451

\bibitem{Landau_Lifshitz1987:Hydro_Book}
Landau L, Lifshitz E. {\it Fluid dynamics}, vol. 6. 2nd ed. Oxford: Reed-Elsevier; 1987.

\bibitem{HuangSatPhys}
K. Huang, {\it Statistical Mechanics}, 2Ed (Wiley, 1987).

\bibitem{Lion2012:JPC}
T. W. Lion, R. J. Allen, {\it Computing the local pressure in molecular dynamics simulations}, J. Phys.: Condens. Matter {\bf 24}, 284133 (2012).

\bibitem{Chandler:book}
D. Chandler, {\it Introduction to modern statistical}, Mechanics. Oxford University Press, Oxford, UK (1987).

\bibitem{AllenTildesley:book}
M. P. Allen, D. J. Tildesley,
{\it Computer Simulation of Liquids}, Oxford University Press (1989).

\bibitem{LandauLifshitz:SatPhysbook}
L.D. Landau, and E.M. Lifshitz, Statistical Physics. Part 1, third ed., Pergamon Press, Oxford, UK (1980).

\bibitem{Sedov1946:PMM}
Sedov LI. Prikl Mat Mek (SSSR). {\bf 10}, 241 (1946).

\bibitem{vonNeumann1947:Book}
von Neumann J. In: Fuchs K, Hirschfelder JO, Magee JL, Peierls R, von Neumann J, editors. The point source solution. Blast wave: Los Alamos; 1947.

\bibitem{Zeldovich1996:Book}
Zel’dovich Y, Raiser Y. Physics of shock waves and high-temperature hydrodynamic phenomena, vol I. New York: Oxford; 1996.

\bibitem{Abolfath2023:YouTube}
R. Abolfath, Lipid Bilayer destruction by ionizing radiation, TS's and shock waves, available online:
https://www.youtube.com/watch?v=0GYgT8sIpUk

\bibitem{Abolfath2023_2:YouTube}
R. Abolfath, Nano-bubbles / cavities induced by charged particles simulated by MD-ReaxFF with $\approx$ 150,000 water molecules. The running time up to 2 ps shows the cavity expands up to 10 nm in diameter. The video is available online:
https://www.youtube.com/watch?v=Sh$\_$6s34WzNE

\bibitem{Ljunggren1997:CS}
S. Ljunggren, J. C. Eriksson,
{\it The lifetime of a colloid-sized gas bubble in water and the cause of the hydrophobic attraction}
Colloids and Surfaces A: Physicochemical and Engineering Aspects {\bf 129-130}, 151-155 (1997).


\bibitem{YouTube_two_SWs}
R. Abolfath, Molecular crowding, and inter-track interaction
The video is available online: https://www.youtube.com/watch?v=lTGNbH2npME

\bibitem{NIST:HeatCapacity}
https://webbook.nist.gov/cgi/cbook.cgi?ID=C7732185\\ \&Units=CAL\&Type=JANAFL\&Table=on

\bibitem{Singh1980:PIAS}
V. P. Singh, A. K. Madan, H. R. Suneja, D. Chand,
{\it Propagation of spherical shock waves in water},
Prec. Indian Acad. Sci. (Engg. Sei.) {\bf 3}, 169-175 (1980).

\bibitem{Jackson1999:book}
J. D. Jackson, {\it Classical Electrodynamics}, Wiley Third Edition (1999).

\bibitem{Ramin_Martin:Unpublished}
R. Abolfath {\em et al.}, M. R\"adler {\em et al.} (unpublished).

\bibitem{Abolfath2019:EPJD}
Abolfath, R.; Helo, Y.; Bronk, L.; Carabe, A.; Grosshans, D.; Mohan, R. {Renormalization of radiobiological response functions by energy loss fluctuations and complexities in chromosome aberration induction: Deactivation theory for proton therapy from cells to tumor control}. \emph{Eur. Phys. J. D} {\bf 2019}, {\it 73}, 64.
%, DOI: 10.1140/epjd/e2019-90263-5

\bibitem{Lamoreaux2005:RPP}
S. Lamoreaux, {\it The Casimir force: Background, experiments, and applications}. Reports on Progress in Physics. {\bf 68}, 201–236 (2005). %doi:10.1088/0034-4885/68/1/r04. S2CID 21131414

\bibitem{Yamamoto2021:RPT}
S. Yamamoto, {\it Discovery of the luminescence of water during irradiation of radiation at a lower energy than the Cherenkov light threshold}, Radiol. Phys. Technol. {\bf 14}, 16–24 (2021). %https://doi.org/10.1007/s12194-020-00588-x

\bibitem{Zheng2020:EES}
H. Zheng, T. Yu, C. Qu, W. Li, Y. Wang,
{\it Basic Characteristics and Application Progress of Supercritical Water}, Earth and Environmental Science {\bf 555}, 012036 (2020).
%doi:10.1088/1755-1315/555/1/012036

%%%%%%%%%%%%%%%%%%%%%%%%%%%%%%%%%%%%%%%%%%%%%%%%%%%%%%%%%%%%%%%%%%%%%%%%%%%%%%%%%%%%%%%%%%%%%%%%%%%%%%%%%%%%%%%%%%%%%%



\end{thebibliography}
\end{document}